\documentclass[12pt]{iopart}

\usepackage{hyperref}
\usepackage{epsfig}
\usepackage{iopams}
\usepackage{graphicx}
\usepackage{dcolumn}
\usepackage{bm}
\usepackage[francais]{babel}
\usepackage{psfrag}

\begin{document}

\title[Imaging the stick-slip peeling of an adhesive tape]
{Imaging the stick-slip peeling of an adhesive tape under a
constant load}

\author{Pierre-Philippe Cortet$^1$, Matteo Ciccotti$^2$ and Lo\"{i}c Vanel$^1$}

\address{$^1$Laboratoire de physique, CNRS UMR 5672,
  Ecole Normale Sup\'erieure de Lyon,
  46 all\'ee d'Italie,
  69364 Lyon Cedex 07, France}

\address{$^2$Laboratoire des Collo\"{i}des, Verres et
    Nanomat\'{e}riaux, CNRS UMR 5587, Universit\'{e} de Montpellier
    II, Place Bataillon, 34095 Montpellier Cedex 5, France}

\begin{abstract}
Using a high speed camera, we study the peeling dynamics of an
adhesive tape under a constant load with a special focus on the
so-called stick-slip regime of the peeling. It is the first time
that the very fast motion of the peeling point is imaged. The
speed of the camera, up to $16000$ fps, allows us to observe and
quantify the details of the peeling point motion during the stick
and slip phases: stick and slip velocities, durations and
amplitudes. First, in contrast with previous observations, the
stick-slip regime appears to be only transient in the force
controlled peeling. Additionally, we discover that the stick and
slip phases have similar durations and that at high mean peeling
velocity, the slip phase actually lasts longer than the stick
phase. Depending on the mean peeling velocity, we also observe
that the velocity change between stick and slip phase ranges from
a rather sudden to a smooth transition. These new observations can
help to discriminate between the various assumptions used in
theoretical models for describing the complex peeling of an
adhesive tape. The present imaging technique opens the door for an
extensive study of the velocity controlled stick-slip peeling of
an adhesive tape that will allow to understand the statistical
complexity of the stick-slip in a stationary case.
\end{abstract}


\maketitle

\section{Introduction}

The peeling dynamics of adhesive tapes and especially its stick-slip
complex regime has been the center of extensive investigations in
recent years. The interest in such a phenomenon has two main
justifications. Industrial processing often requires to peel ribbon
of different kinds at very high velocities. In this situation, the
stick-slip phenomenon and its jerky dynamics can cause important
problems including delays on the assembly line. Additionally,
understanding the physics of the adhesive tape peeling is precious
for modeling and testing the resistance of elastomer-substrate
joints, and this seemingly simple system allows to gain deeper
insights into more subtle aspects of the physics of adhesion
\cite{RYSHE1996}. Despite extensive experimental and modeling
efforts, our understanding of the jerky behavior experienced in the
so-called stick-slip regime is still limited. The phenomenon remains
highly non-linear and the dynamics shows a variety of instabilities
and structures \cite{CICCO2004}. Several dynamical models were
developed with an increasing degree of realism, leading to an
increasing complex dynamics
\cite{HONG1995,CICCO1998,ANANT2004,ANANT2005,ANANT2006}. Each model
is characterized by a series of seemingly trivial assumptions which
progressively revealed to have a crucial effect on the dynamical
aspects of the problem.

Due to the very rapid nature of the stick-slip dynamics, the main
physical quantity generally accessed to in the experiments is the
distribution of the time intervals between successive events that
can be identified through the measurement of the related acoustic
or photonic emissions \cite{CICCO2004,BARQU1997}. In such
experiments, an increasing irregularity of the stick-slip dynamics
has been observed for an increasing peeling velocity. This
behavior reveals to be more rich and complicated than a simple
bifurcation route to chaos. In particular, one can observe
hierarchical structures in a definite traction velocity range that
can suggest the emergence of complexity.

Even though a lot of experimental studies of adhesive tape peeling
has been performed, the direct observation of the local motion of
the peeling point in the stick-slip regime has not been done for
the moment. In this paper, we present an experimental procedure
for imaging directly, using a high speed camera, the dynamics at
the peeling point, especially during the stick-slip phase (cf.
\href{video/video1.avi}{video 1}). We test this technique in
experiments where a constant load is applied to the tape. These
very first experiments showing the true stick-slip dynamics of the
peeling point in an adhesive tape already give very precise
informations that will help to screen among the various
assumptions usually done in theoretical models. The technique
opens the way for many more experimental investigations that
should clarify the physics at stake in the peeling of an adhesive
tape.

\section{\label{previous}Previous experiments of adhesive tape peeling}

The experiments on the peeling of an adhesive tape are generally
performed using two different setups. In the first one, the
peeling is studied when a constant traction velocity $V$ is
imposed onto the free end of the tape by the action of an electric
motor. In this case, with a fixed geometry, $V$ is the only
dynamical control parameter. In a second type of experiment the
peeling is studied when a constant applied load is clamped to the
tape free end and the control parameter is the imposed force $F$.
In these experiments, the limit between the adhesive tape ribbon
and the free tape may be seen as a crack tip propagating at a
speed $v$.

\begin{figure}
\psfrag{Y}[c]{$\log f$} \psfrag{X}[c]{$\log v$}
\psfrag{V1}[c][][0.8]{$V_B$} \psfrag{V2}[c][][0.8]{$V_D$}
\psfrag{F1}[c][][0.8]{$F_B$} \psfrag{F2}[c][][0.8]{$F_D$}
    \centerline{\includegraphics[width=9cm]{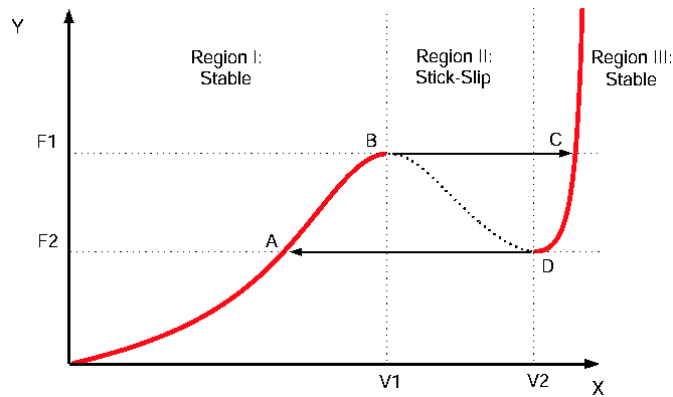}}
    \caption{\label{SchemaFV} Schematic relation between the peeling
    force $f$ and the peeling velocity $v$ at the peeling crack line.
    These variables refer to the local dynamics of the peeling point
    and correspond to the tensile force $F$ and velocity $V$ at the free end of the
    tape only when the peeling is regular and the peeling angle $90\degres$.
    The sigmoidal shape is responsible for the hysteretic behavior and
    therefore for the stick-slip dynamics.}
\end{figure}

It is widely admitted in the literature
\cite{CICCO2004,HONG1995,CICCO1998,ANANT2004, ANANT2005,
ANANT2006,BARQU1997} that there is a fundamental relation between
the local peeling force $f$ and the peeling velocity $v$ at the
peeling point ($f$ is equal to the tensile force $F$ at the free
tape end in an idealized case where the peeling angle is
$90\degres$ and the peeling is regular). A schematic
representation of such relation is plotted in figure
\ref{SchemaFV}. The sigmoidal shape of this curve, that is a
consequence of the adhesive material rheology, presents three
remarkable regions: two stable ones (I and III) and an unstable
one (II). A full theoretical understanding of the shape of this
curve is not available for the moment, but it has been known, at
least since Prandtl \cite{PRAND1928}, that such a dependence will
lead to a hysteretic behavior of the peeling point dynamics. A
similar hysteresis of the force-velocity response is proposed to
explain the pinning and depinning dynamics of charge-density
waves, vortex arrays in semi-conductors \cite{VINOK1997,MARCH2000}
and possibly the stick-slip behavior of contact lines
\cite{SCHAE2000} or magnetic domain walls \cite{URBAC1995}.

Barquins and Maugis \cite{BARQU1986,MAUGI1988} performed a series of
experiments at constant traction velocity $V$. The observed dynamics
exhibits the following behavior: at slow traction velocity, the tape
is peeled regularly and the dynamics is stationary (branch I); at
high traction velocity, the dynamics is also regular, but much more
rapid (branch III); in the intermediate range of $V$, a stick-slip
phenomenon appears, the peeling of the tape being jerky with
emission of a characteristic noise. In this regime, for increasing
values of the traction velocity (from $V_B$ to $V_D$), the
stick-slip motion is at first rather periodic, then it becomes more
and more irregular. It has been argued that the irregular motion
corresponds to chaotic orbits \cite{HONG1995}. In this regime, the
local dynamics at the peeling point is expected to follow a complex
trajectory, still not experimentally resolved, around the hysteretic
region of the $f-v$ curve.

\begin{figure}
    \centerline{\includegraphics[width=9cm]{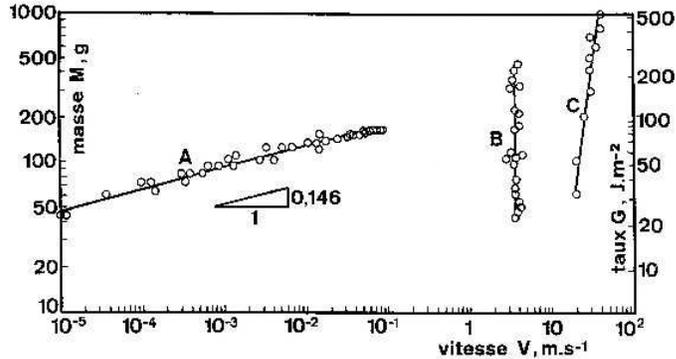}}
    \caption{\label{GrafVP} Applied mass as a function of the mean mass
    falling velocity $\langle v \rangle$ as reported in \cite{BARQU1997}.}
\end{figure}

In contrast, when a constant force $f$ is applied to the peeling
point, a stable regime always exists for the peeling (cf. figure
\ref{SchemaFV}). However, for an applied force between $F_D$ and
$F_B$, there are two stable solutions, one on branch I (AB) and
the other on branch III (DC) (cf. figure \ref{SchemaFV}). In
experiments where the peeling was produced by a constant applied
force $F$, with the help of a set of different dead loads clamped
to the extremity of the free tape \cite{BARQU1997}, several
different regimes were observed for a given load depending on the
way the experiment is started. The simplest experiment consists in
applying the load instantaneously for an initial zero velocity of
the peeling. In this case, where the initial condition of the
peeling is out of equilibrium, the system reaches, for loads under
$F_B$, a stable and regular peeling regime corresponding to branch
I (branch A in figure \ref{GrafVP}). Forcing the peeling with a
large enough initial velocity, it is as well possible to observe
the stable and regular regime corresponding to branch III (branch
C in figure \ref{GrafVP}) for loads over $F_D$. Moreover, an
unexpected stick-slip regime was observed between the two stable
branches (AB) and (DC), for loads between $F_D$ and $F_B$, when
introducing a moderate initial velocity to the peeling. The
existence of this regime shows the metastability of the (AB) and
(DC) branches and was attributed to the inertia of the falling
load that can not maintain instantaneously a constant force at the
peeling point. In this regime, the local force and velocity at the
peeling point are following cycles in the (ABCD) region of figure
\ref{SchemaFV}. Experimentally, it was observed that the time
averaged value $\langle v \rangle$ of the peeling velocity in this
stick-slip regime reaches a constant value that is almost
unsensitive to the load over one order of magnitude (cf. figure
\ref{GrafVP}). Finally, for loads over the critical load $F_B$
(and typically less than $3 F_B$) and an initial zero velocity of
the peeling, a stick-slip regime arising spontaneously can be
observed. The characteristics of this stick-slip regime are
totally consistent with the one observed for lower loads.

\section{\label{expsetup} The experimental setup}

\begin{figure}
\psfrag{R}[c]{$R$} \psfrag{w}[c]{$\omega$} \psfrag{T}[c]{$\theta$}
\psfrag{a}[c][][0.9]{$\alpha$} \psfrag{b}[c][][0.9]{$\beta$}
\psfrag{M}[l]{$m$}
    \centerline{\includegraphics[height=9cm]{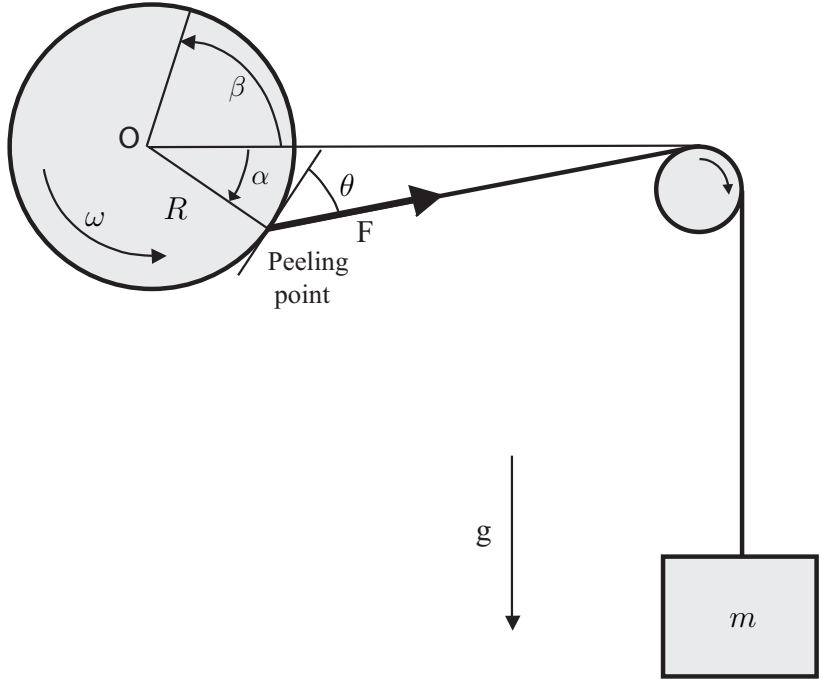}}
    \caption{\label{Schema} Experimental setup and variables. The
    angles $\alpha$ and $\beta$ are algebraic and oriented trigonometrically.
    Roller radius: $5.85$cm$>2R>3.65$cm, roller and tape width:$1.95$cm, tape thickness:$50\mu$m.}
\end{figure}

The results presented in this paper refer to experiments where an
adhesive roller tape (3M Scotch\textsuperscript{\textregistered}
$600$) is peeled off by applying a constant load. Actually, we
attached a mass to the tape extremity and let it fall to the floor
from a height of about $1.6$m with the roller mounted on a pulley
rotating freely (cf. figure \ref{Schema}). There is an additional
pulley, between the roller and the mass, over which the
non-adhesive side of the tape rolls. The distance between the
pulley and the roller is $0.80$m. The adhesive tape and the basic
loading scheme that have been used here are of the same kind as in
\cite{BARQU1997}. In our experiment, we study the transient
response of the peeling of the adhesive tape when a constant load
is applied. More precisely, we introduce, depending on the load,
different initial peeling velocities in order to enter the
stick-slip regime during the fall.

\begin{figure}
    \centerline{\includegraphics[width=10cm]{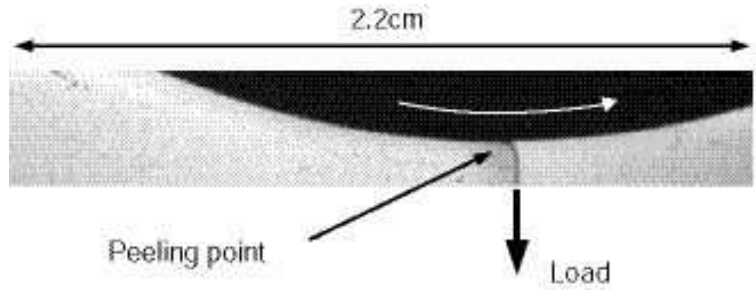}}
    \caption{\label{image1} Image of the region near the peeling point ($512*64$ pixel$^2$).}
\end{figure}

The local dynamics of the peeling point has been imaged using a
high speed camera (Photron Ultima $1024$) at a rate of $8000$ or
$16000$fps. The camera provides a $1024*1024$pixel$^2$ resolution
when used at low frame rates. However, as the frame rate is
increased, resolution is reduced. We get a very elongated image of
$512*64$ (resp. $256*32$) pixel$^2$ at the frame rate of $8000$fps
(resp. $16000$fps). The elongated shape of the images leads us to
focus on the peeling point region (cf. figure \ref{image1}). The
longest direction of the image has been set perpendicular to the
pulling direction of the applied load so as to maximize the
resolution of the peeling point motion. One can see in figure
\ref{image1} a typical image showing the peeling point, a part of
the rotating roller and the beginning of the free tape. On the
background, one can see defects of different sizes and shapes that
actually are deposited on a transparent film that is attached to
the tape roller. The rotational speed of these defects is
therefore the same as the roller's one.

The recording of each movie is synchronized with the arrival of
the load on the ground level using a mechanical switch that
generates a trigger signal when the falling load hits it. The
camera works in a ``trigger end'' mode, i.e. it keeps acquiring
images until receiving the trigger signal. Consequently, in our
movies, the last image corresponds to the moment when the load
reaches the ground level.

\section{\label{extraction}The extraction of the peeling dynamics from the movies}

The movies that have been recorded (cf.
\href{video/video1.avi}{video 1}) allow to access the curvilinear
position of the peeling point $\ell_{\alpha}= R \, \alpha$ as well
as the curvilinear position of the scotch roller $\ell_{\beta}=R\,
\beta$ in the laboratory reference frame as a function of time
($\alpha$ measures the angular position of the peeling point,
$\beta$ the rotation of the roller and $R$\footnote{$R$ is
obviously a function of time during a peeling experiment.
Therefore, before each experiment, we measure  the value of $R$.
During one experiment, we neglect the variations of $R$ that are
less than $3\%$.} is the roller radius (cf. figure \ref{Schema})).
Once we know the position of the peeling point and the rotation of
the roller in the laboratory reference frame, we can easily
compute the curvilinear position of the peeling point
$\ell_{\gamma}=\ell_{\alpha}-\ell_{\beta}$ in the roller reference
frame and the corresponding angular position
$\gamma=\alpha-\beta$.

\begin{figure}
    \centerline{\includegraphics[width=10cm]{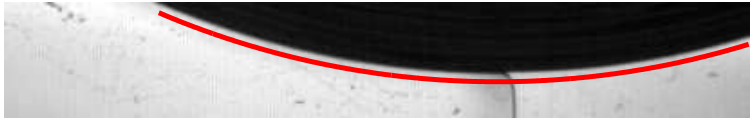}}
    \caption{\label{image2} Image of the region near the peeling
    point (same one as in figure \ref{image1}) and the extracted pixels line on the circular shape.}
\end{figure}

\begin{figure}
\psfrag{A}{(a)} \psfrag{B}{(b)}
    \centerline{\includegraphics[width=11cm]{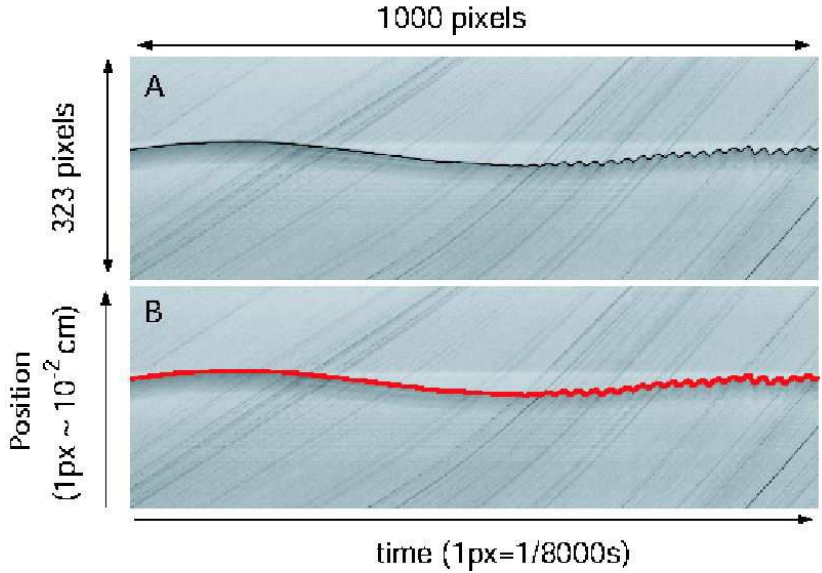}}
    \caption{\label{spatiotemp} (a) Spatiotemporal image of the peeling
    point region. (b) Same image with superimposed the extracted position signal.}
\end{figure}

In order to access $\ell_{\alpha}$ and $\ell_{\beta}$ as a
function of time, we extract for each image in a movie a line of
pixels (solid line in figure \ref{image2}) that follows the
circular shape of the tape roller surface at a distance of a few
pixels from this surface. This line of pixels is darker in a zone
corresponding to the peeled ribbon. The pixel position of this
zone gives with an analytical correction\footnote{The angular
position  $\ell_{\alpha}$ is related to the pixel position
$n_{\rm{\small px}}$ in the extracted pixels line as follow:
$\ell_{\alpha}=R\, \arcsin\left(\frac{n_{\rm{\small
px}}-n^o_{\rm{\small px}}}{R}\right)$ where $n^o_{\rm{\small px}}$
is the position of the roller axis.} the angular position of the
peeled tape near the peeling point. Building an image with such an
extracted pixels line for each time step of the movie leads to a
spatiotemporal representation of the position of the peeling point
as we can see on figure \ref{spatiotemp}(a).

In such an image, we can extract the pixel position of the peeling
point at each time step (cf. figure \ref{spatiotemp}(b)) and then
have access to the full time-resolved curve of the peeling point
position in the laboratory reference frame $\ell_{\alpha}$. We do
this by correlating the typical grey level profile around the
peeling point with each pixels line. We have used the gray-level
information in order to get sub-pixel accuracy on $\ell_{\alpha}$.

The tilted dark lines, that one can see in the background of
figure \ref{spatiotemp}, correspond to the motion of the defects
on the film that is attached to the roller. Their local slope
represents the local rotational velocity of the roller. In order
to have access to a full time-resolved motion of the roller, we
actually built an image correlation technique that uses these
defect lines. To improve the spatial resolution of the correlation
technique, we have again used the gray-level information in order
to get sub-pixel accuracy. In this manner, we are able to access
the velocity $\dot{\ell}_{\beta}$ of the roller's rotation with an
excellent precision. From the velocity signal, we can easily
compute the roller position $\ell_{\beta}$ through numerical
integration. A consistency check is performed by comparing the
integrated position to the actual length of the peeled ribbon when
the load reaches the ground.

\section{Dependance of the peeling dynamics with the applied load and initial velocity}\label{VP}

In our experiment, when applying a load to the tape extremity and
starting from an initial zero peeling velocity, the peeling is
regular and quickly reaches a stationary regime (cf.
\href{video/video2.avi}{video 2}) for applied masses below a
critical mass, $m_B =(235 \pm 5)$g, with a slight dependence on
the tape sample. The system quickly reaches equilibrium in the
region I of figure \ref{SchemaFV} and the critical mass actually
corresponds to the critical load $F_B$. In this regime, the mass
falling velocity, that is equal to the average roller rotational
velocity $\langle \dot{\ell}_{\beta} \rangle$, appears to be
fairly constant during the fall and increases with the applied
mass (for $m<m_B$) up to a value $v_{c}=(0.20 \pm
0.03)$m.s$^{-1}$. When $m$ becomes larger than $m_B$, a stick-slip
peeling dynamics appears spontaneously during the fall (cf.
\href{video/video1.avi}{video 1}) along with a characteristic
acoustic emission. Coincidentally, the mean velocity of the entire
mass fall jumps to higher values. The initial condition in these
experiments is out of equilibrium since it starts on the left end
of the $v$-axis away from the $f-v$ curve (cf. figure
\ref{SchemaFV}). As it has been noticed in section \ref{previous},
it is also possible to trigger a stick-slip dynamics for $m<m_B$
by introducing manually an initial velocity to the peeling.

In the next sections, we will present a detailed analysis of the
transient peeling dynamics in the two cases where stick-slip has
been heard: $m<m_B$ with an initial velocity, $m>m_B$ without
initial velocity. From now on, these two regimes will be referred
to respectively as the triggered and spontaneous stick-slip
regimes.

\section{The rotation dynamics of the roller}\label{rotreflab}

\subsection{Triggered case}

\begin{figure}[h!]
\psfrag{X}[][][0.9]{$t$ (s)}
\psfrag{Y}[c][][0.9]{$\dot{\ell_{\beta}}$ (m.s$^{-1}$)}
\psfrag{Z}[c][][0.9]{$\ddot{\ell_{\beta}}$ (m.s$^{-2}$)}
\psfrag{A}[c][][0.9]{(a)} \psfrag{B}[c][][0.9]{(b)}
\psfrag{C}[c][][0.9]{(c)} \psfrag{D}[c][][0.9]{(d)}
\psfrag{E}[c][][0.9]{(e)} \psfrag{F}[c][][0.9]{(f)}
    \centerline{\includegraphics[width=7cm]{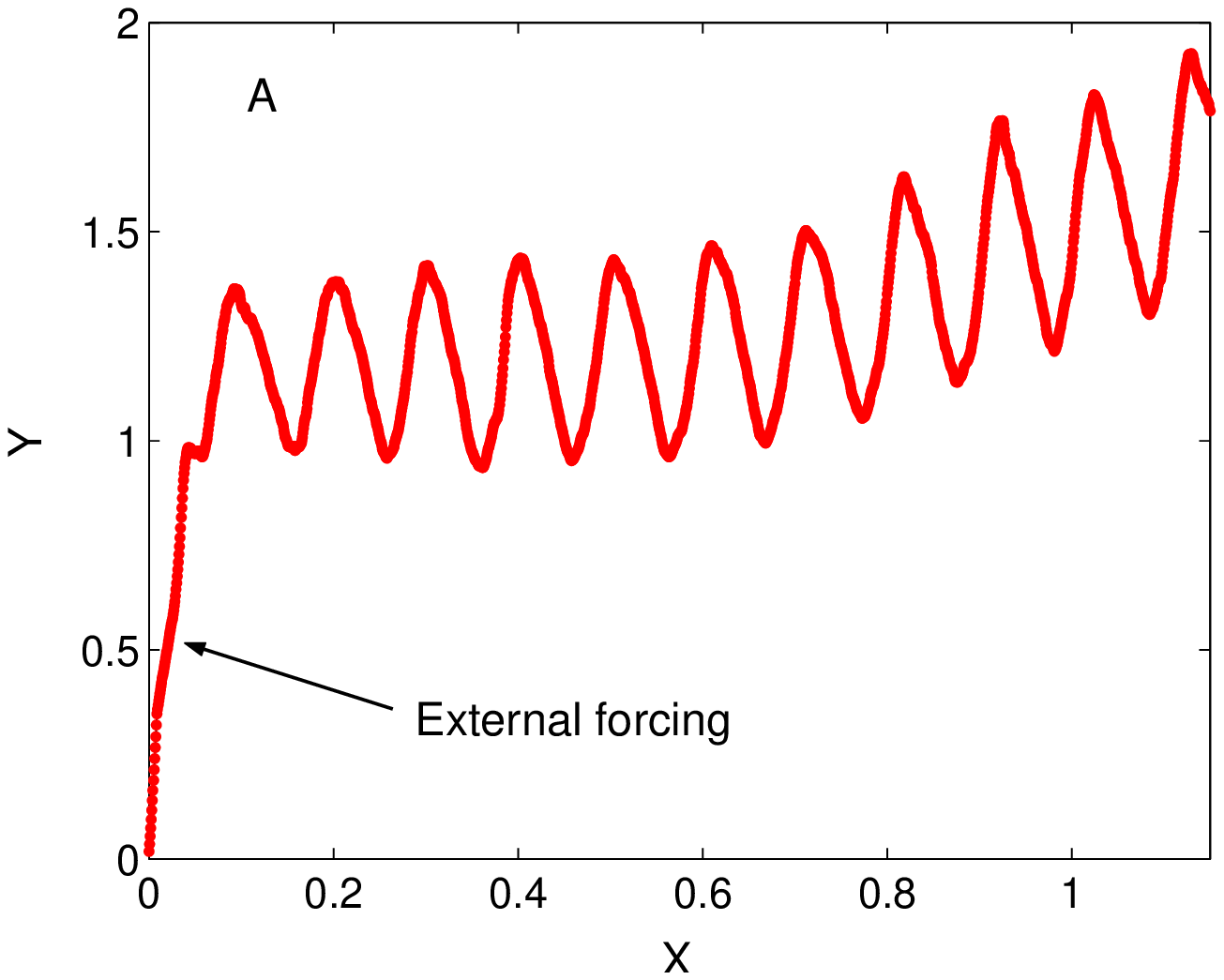}
    \includegraphics[width=7cm]{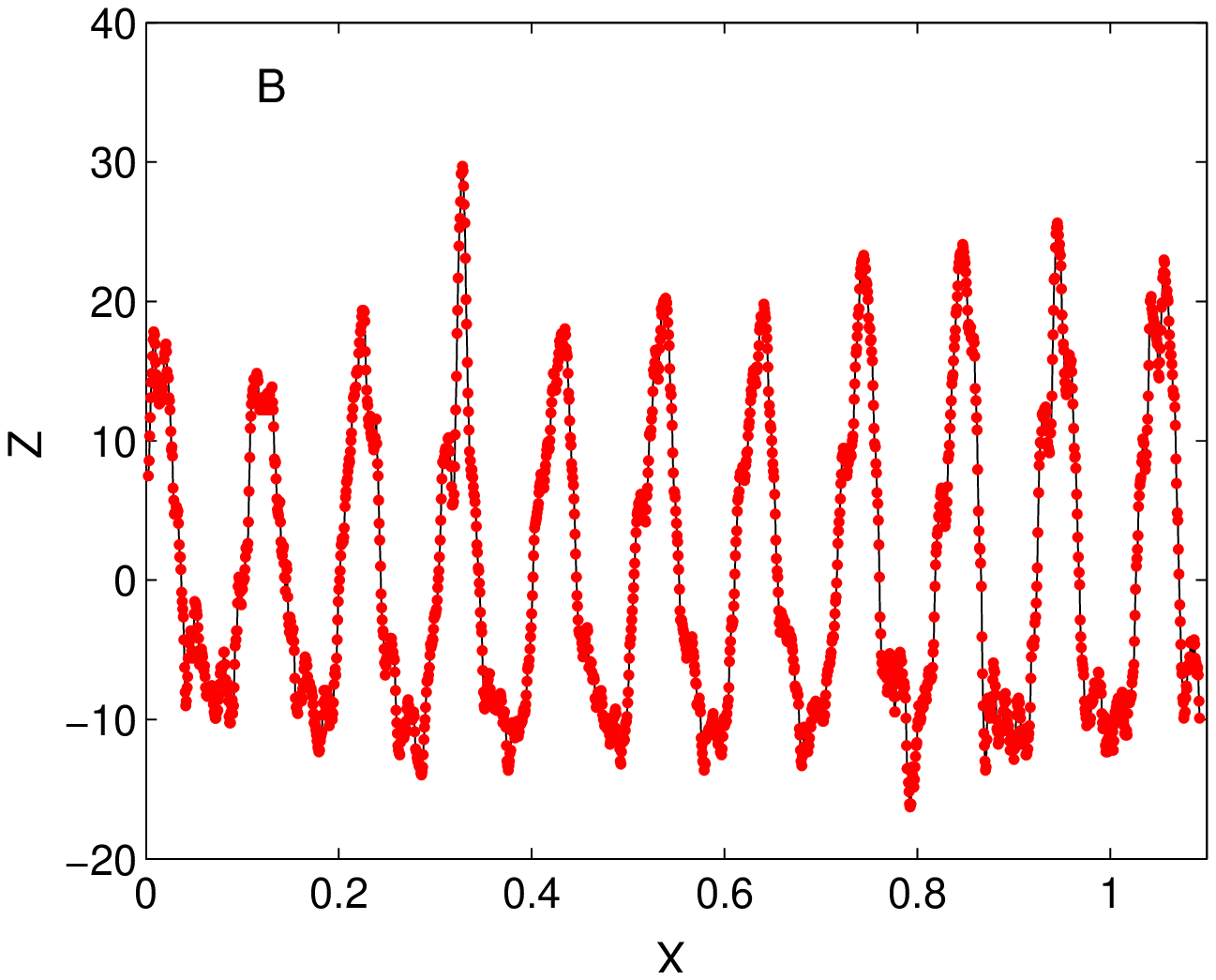}}
    \vspace{0.2cm}
    \centerline{\includegraphics[width=7cm]{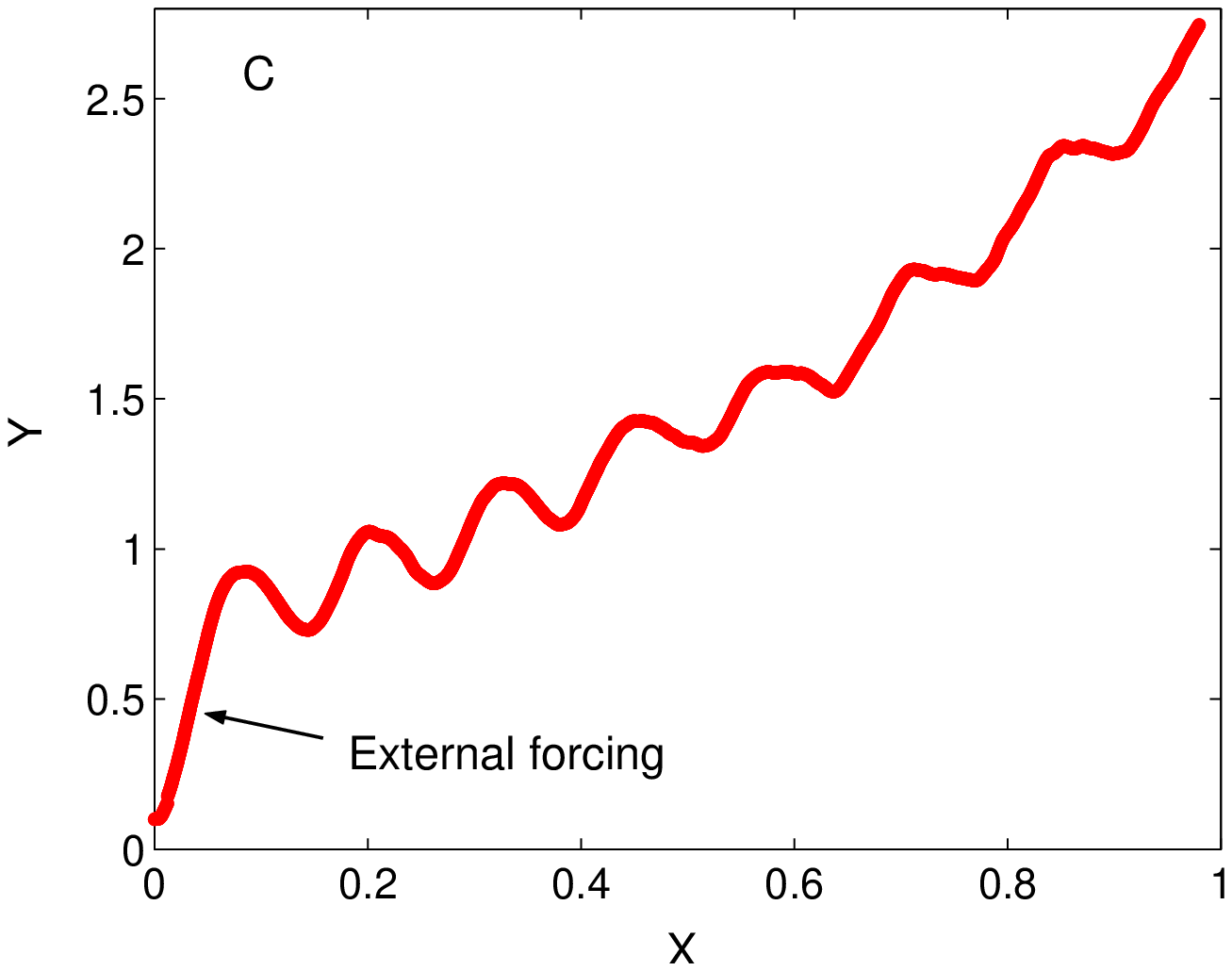}
    \includegraphics[width=7cm]{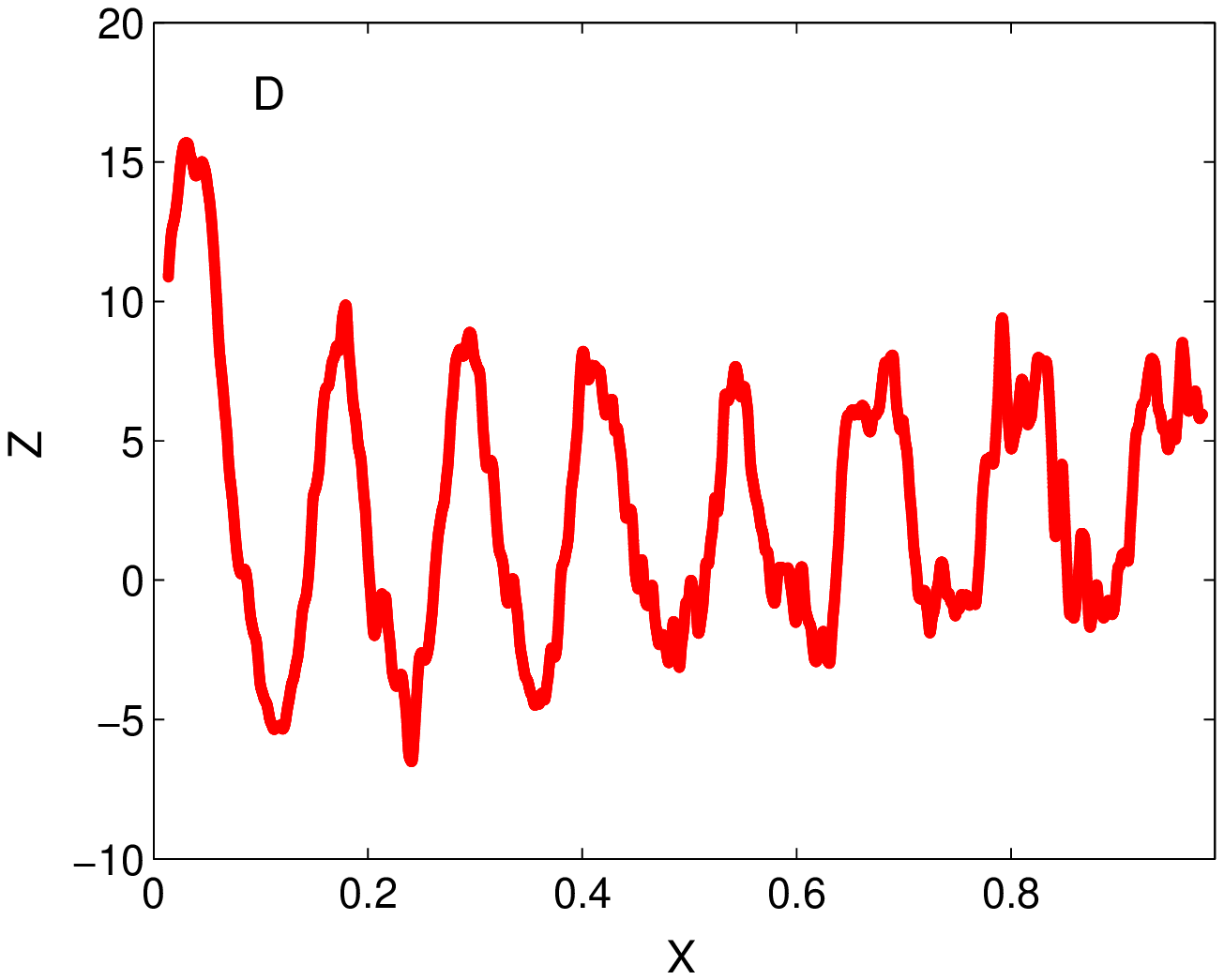}}
    \vspace{0.2cm}
    \centerline{\includegraphics[width=6.95cm]{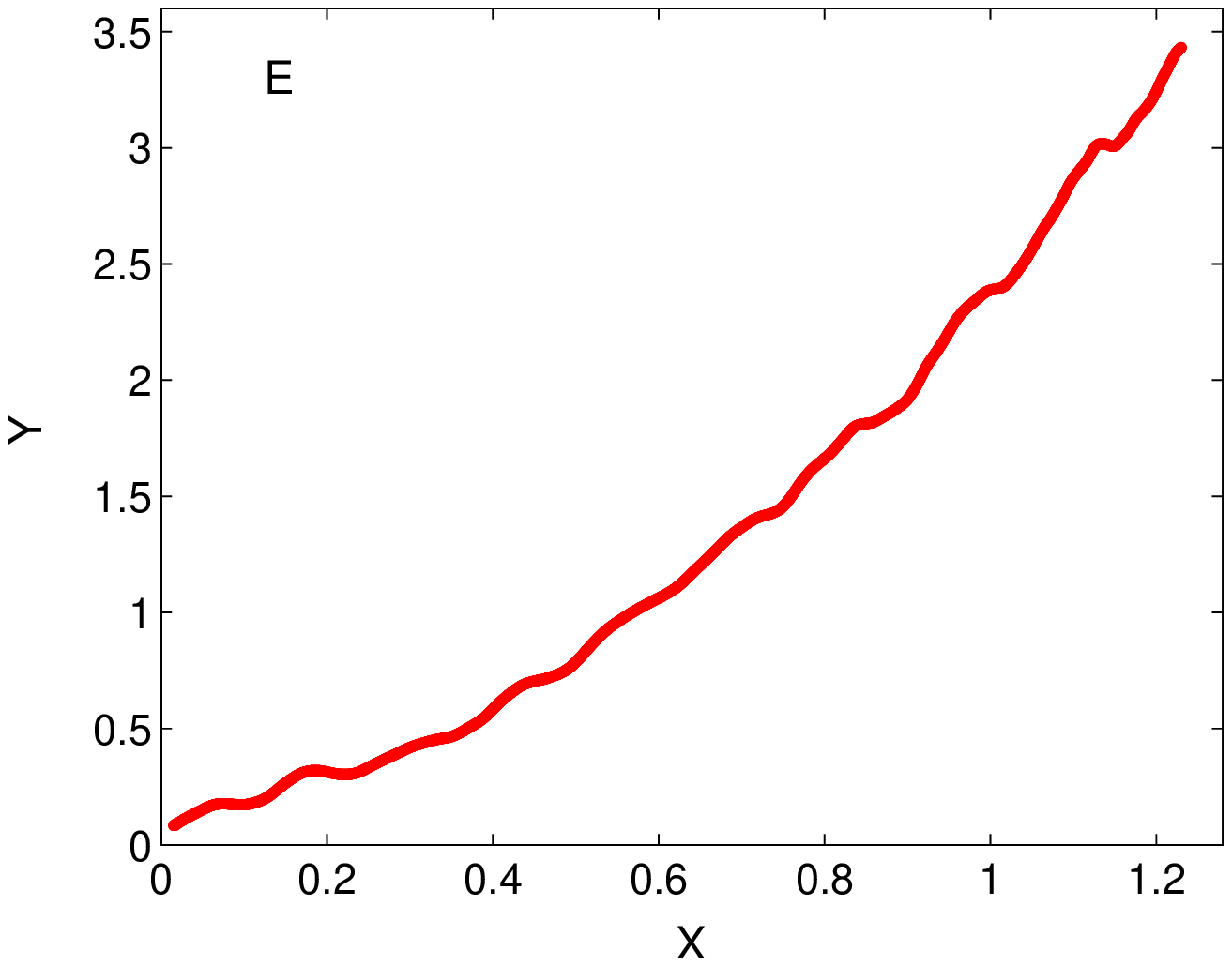}
    \includegraphics[width=6.9cm]{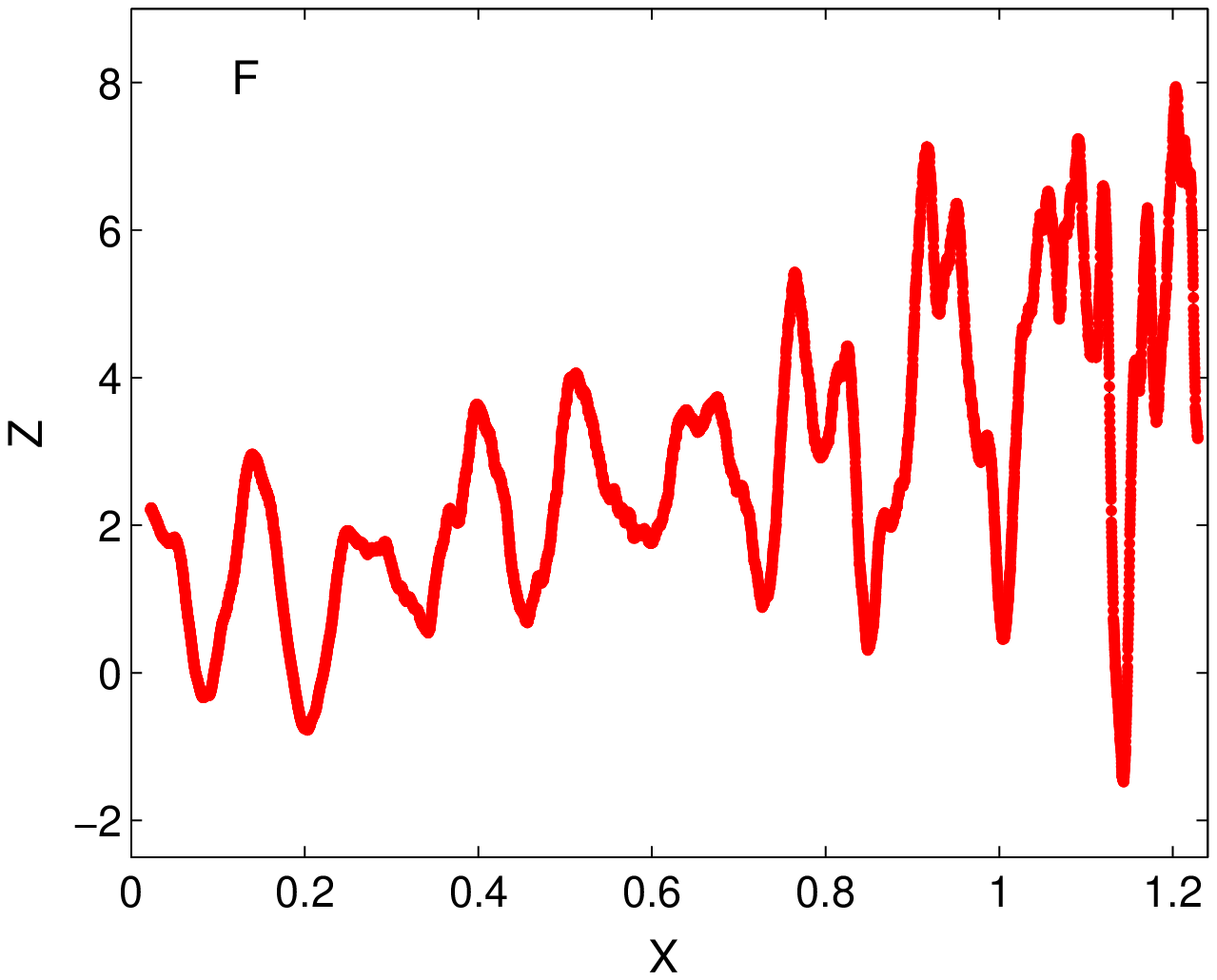}}
    \caption{\label{vitrotsub170} (a), (c) and (e), rotation velocity
    $\dot{\ell_{\beta}}$, (b), (d) and (f), corresponding (respectively to (a), (c) and (e))
    acceleration $\ddot{\ell_{\beta}}$,
    as a function of the time. (a), (b), (c) and (d) curves correspond to a triggered stick-slip peeling
    experiment performed with $m=170$g ((a) and (b) curves) and with $m=195$g ((c) and (d) curves).
    (e) and (f) curves correspond to a spontaneous stick-slip peeling
    experiment performed with $m=245$g.}
\end{figure}

In figure \ref{vitrotsub170}(a), we present the rotation velocity
$\dot{\ell_{\beta}}$ of the roller as a function of time for a
typical experiment performed with $m=170$g$<m_B$ in the case where
the stick-slip has been triggered. First, we notice the initial
acceleration of the rotation, with $\dot{\ell_{\beta}}$ getting
from zero to about $1$m.s$^{-1}$, that corresponds to the manual
external forcing. Next, we see an oscillation of the rotation
velocity at a frequency of about $(9.8\pm0.1)$Hz and an amplitude
of $(0.50\pm 0.05)$m.s$^{-1}$. The amplitude of these oscillations
is large. In figure \ref{vitrotsub170}(b), we present the
corresponding acceleration of the roller $\ddot{\ell_{\beta}}$. In
this figure, we see that the acceleration oscillates with time
between $(-11\pm 2)$m.s$^{-2}$ and $(21\pm 2)$m.s$^{-2}$.

The acceleration oscillates around a non-zero average value (mean
acceleration over an oscillation) of about $0.74$m.s$^{-2}$ over
the whole fall of the mass. This leads to an increase in the
average rotation velocity from $1.15$m.s$^{-1}$ to
$1.65$m.s$^{-1}$. This increase is related to an acceleration of
the average peeling velocity $\langle v \rangle$ during the fall
of the mass (cf. section \ref{peelrefrol}). We will see in the
next section that the stick-slip heard during the fall of the mass
has completely disappeared at the end of the fall and is therefore
only a transient phenomenon.

As the applied mass is increased, we observe that the amplitude of
the oscillations for the velocity and the acceleration decreases
while the mean acceleration increases ($m=195$g$<m_B$ in figures
\ref{vitrotsub170}(c) and (d)).

\subsection{Spontaneous case}

In figure \ref{vitrotsub170}(e), we present the rotation velocity
$\dot{\ell_{\beta}}$ of the roller as a function of time for a
typical experiment performed with $m=245$g$>m_B$ in the case where
the stick-slip is spontaneous. As well as in the triggered case,
we see an oscillation of the rotation velocity at a frequency of
about $(8.5\pm 0.1)$Hz. The amplitude of this oscillation is
however much smaller than in the triggered case: $(0.015\pm
0.005)$m.s$^{-1}$. In contrast, as one can see on figure
\ref{vitrotsub170}(f), the mean acceleration over an oscillation
of the rotation of the roller during the fall of the mass is
larger and it gradually increases from $1$m.s$^{-2}$ to
$4$m.s$^{-2}$. This non-zero mean acceleration results in an
important increase in the mean velocity from zero to about
$3.5$m.s$^{-1}$. As for the triggered case, the stick-slip reveals
to be only a transient phenomenon in these conditions (cf. section
\ref{peelreflab}).

We should point out that the observed increasing acceleration,
both in the triggered and spontaneous case, is inconsistent with
the observation of an asymptotic constant velocity in the force
controlled stick-slip regime as reported in \cite{BARQU1997}.

\subsection{\label{velosc}The velocity oscillations}

\begin{figure}
    \centerline{\includegraphics[width=9cm]{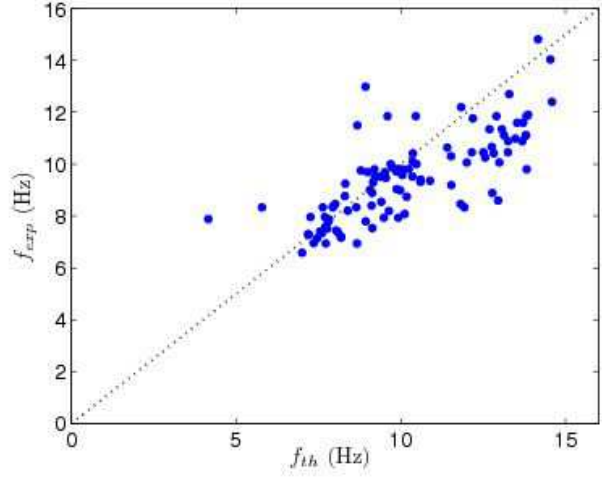}}
    \caption{\label{freq} Experimental oscillation frequency as a
    function of the theoretical prediction (cf. equation \ref{frequency}),
    taking into account the accelerated motion of the load. The data
    corresponds to different applied masses $m=170, 195, 245, 265$g,
    various radius $5.60$cm$>R>3.60$cm and different moments during
    the fall of the mass.}
\end{figure}

The origin of the velocity oscillations can be understood as a
consequence of the interplay between the inertia of the roller and
the moment applied to the roller by the peeling force. Since the
oscillations exist even when there is no stick-slip, it is not
necessary to take into account the stick-slip variations of the
peeling force to explain the oscillations. However, it is
important to take into account the acceleration $a$ of the load
which reduces the tensile force on the peeling point:
$F=m(g-a)$\footnote{It is almost rigorously true that
$a=\ddot{\ell_{\gamma}}$}. Since the length of the peeled tape is
long compared to the radius of the roller, we will assume that the
direction of the tensile force is horizontal (i.e. $\theta \simeq
\pi/2 + \alpha$). The basic equation of motion for the roller is
then:
\begin{equation}
I \ddot{\beta} = - m (g-a) R \sin\alpha
\end{equation}
where $I$ is the moment of inertia of the roller. In this
equation, since it evolves quite slowly in time, we will consider
$a$ to be a constant parameter and estimate it with: $a=\langle
\ddot{\ell_{\gamma}} \rangle_{T}$, where $T$ is the period of the
velocity oscillation. Then, if we neglect the relative motion of
the peeling point with respect to the roller (which is rather fast
with respect to the velocity oscillations), we can write that
$\gamma=\alpha-\beta=0$. Thus, considering only small values of
$\alpha$, the roller's natural oscillation frequency is:
\begin{equation}\label{frequency}
\omega= \sqrt{\frac{m (g-\langle \ddot{\ell_{\gamma}}
\rangle_{T})R}{I}}
\end{equation}
In figure \ref{freq}, we show the experimental oscillation
frequency against the theoretical one $f_{th}=\omega/2\pi$. We
observe that there is a reasonable quantitative agreement between
the two frequencies that confirms the relevance of equation
\ref{frequency} even though a lot of approximations have been
used. Taking into account the motion of the peeling point would
lead to corrections on the predicted frequency that are typically
smaller than the experimental scatter of the data.

\section{The peeling point dynamics}

\subsection{In the laboratory reference frame}\label{peelreflab}

The main concern of this paper is to explore the rapid stick-slip
dynamics of the peeling point and access directly variables that
were previously only guessed through acoustic or photonic
emissions. Consequently, we will in this section have a closer
look at the peeling point motion during the stick-slip regime.

\subsubsection{Triggered case}

\begin{figure}
    \centerline{\includegraphics[width=11cm]{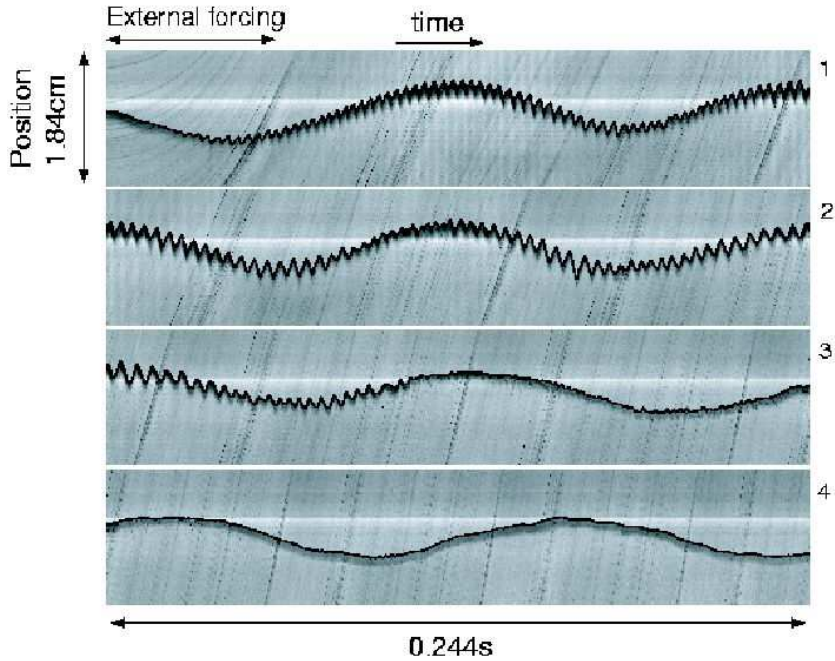}}
    \caption{\label{spatiotemp2} Spatiotemporal image of the peeling
    point region for a triggered stick-slip peeling experiment
    performed with $m=195$g. The extracted peeling point position
    has been added in black.}
\end{figure}

In figure \ref{spatiotemp2}, we present the full spatiotemporal
image of the peeling point region, built according to the process
exposed in section \ref{extraction}, for a typical triggered
stick-slip experiment performed with $m=195$g$<m_B$.

This spatiotemporal image contains a lot of information. First, we
note that there are low frequency oscillations of the peeling
point position. There is a clear correlation between these
oscillations and the one observed for the roller's rotation
velocity. This correlation was expected since the moment on the
roller resulting from the applied tensile force $F$ depends on the
angle $\alpha$, thus on the position of the peeling point in the
laboratory reference frame. Next, during the early stage, we can
notice the huge curvature of the background black lines that
corresponds to the initial acceleration due to external forcing.
The stick-slip actually starts to develop during this phase. Then,
its amplitude grows with time. However, later during the
experiment, the stick-slip amplitude decreases back until a
complete disappearance of the stick-slip motion. This evolution of
the peeling dynamical features occurs while the mean peeling
velocity is increasing. Whether this eventually leads or not the
system to reach an equilibrium state corresponding to branch III
of figure \ref{SchemaFV} is still unclear.

\begin{figure}
\psfrag{X}[][][0.9]{$t$ (s)} \psfrag{Y}[c][][0.9]{$\ell_{\alpha}$
(m)}
    \centerline{\includegraphics[width=11cm]{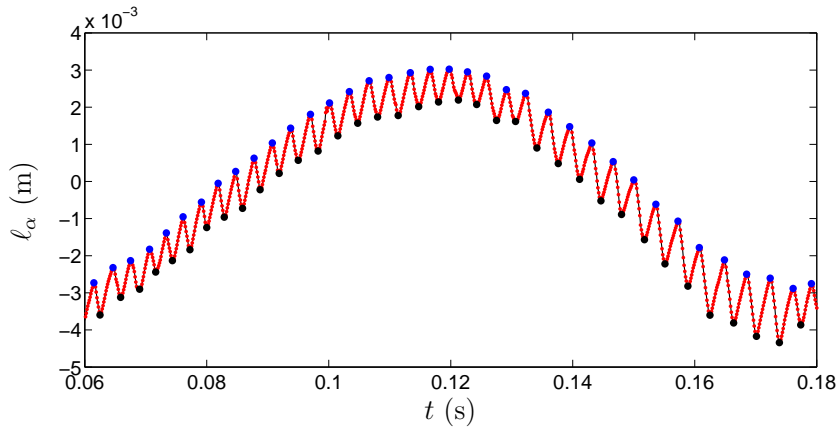}}
    \caption{\label{ppmotion195_03} Position of the peeling point
    in the laboratory reference frame, $\ell_{\alpha}$, as a function
    of time for a triggered stick-slip peeling experiment
    performed with $m=195$g.}
\end{figure}

In figure \ref{ppmotion195_03}, we show part of the stick-slip
motion signal extracted from the spatio-temporal image. The stick
(resp. slip) phase corresponds to an increase (resp. decrease) of
$\ell_\alpha$.  From such a signal, we have been able to study the
evolution of the stick-slip amplitude, durations and velocities
(stick and slip) as a function of time.

\subsubsection{Spontaneous case}

\begin{figure}[h!]
    \centerline{\includegraphics[width=11cm]{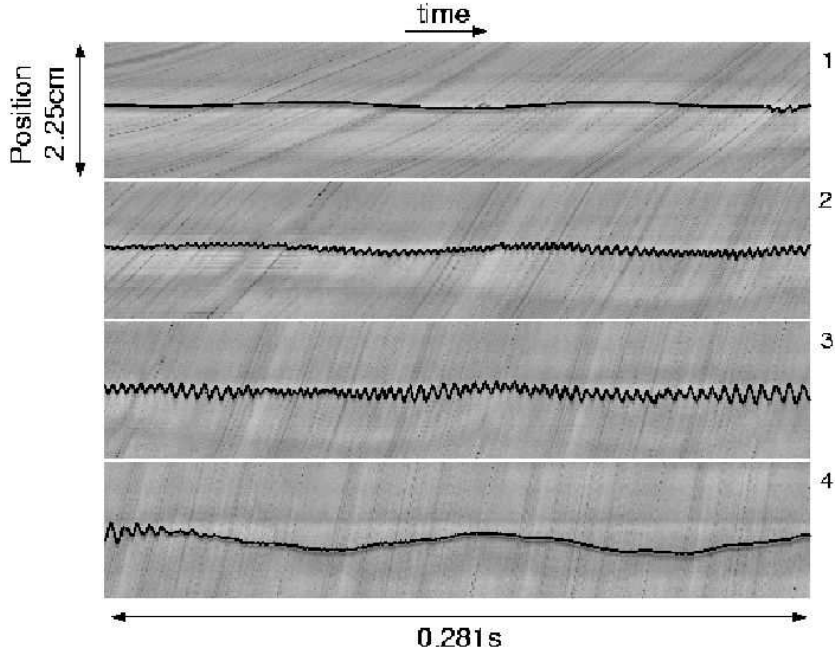}}
    \caption{\label{spatiotemp3} Spatiotemporal image of the peeling
    point region for a spontaneous stick-slip peeling experiment
    performed with $m=245$g. The extracted peeling point position
    has been added in black.}
\end{figure}

In figure \ref{spatiotemp3}, we present the full spatiotemporal
image of the peeling point region for a typical spontaneous
stick-slip experiment performed with $m=245$g$>m_B$.

\begin{figure}[h!]
\psfrag{X}[][][0.9]{$t$ (s)} \psfrag{Y}[c][][0.9]{$\ell_{\alpha}$
(m)}
    \centerline{\includegraphics[width=11cm]{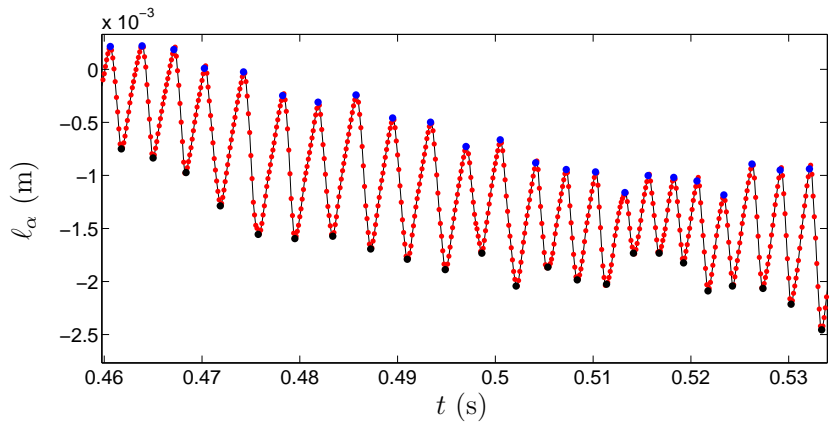}}
    \caption{\label{ppmotion245_02} Position of the peeling point
    in the laboratory reference frame, $\ell_{\alpha}$, as a function
    of time for a spontaneous stick-slip peeling experiment
    performed with $m=245$g.}
\end{figure}

As in the previous paragraph, there is a low frequency oscillation
of the peeling point position, but with a smaller amplitude that
corresponds to the decrease of the roller's oscillations amplitude
with $m$ that has been noticed in the previous section. The
stick-slip peeling is not initially present at the beginning of
the experiment. It starts to develop and grow after a certain time
before its amplitude decreases back until a complete disappearance
of the stick-slip motion. Part of the extracted stick-slip signal
is shown in figure \ref{ppmotion245_02}.

To conclude this subsection, it is important to emphasize that the
stick-slip regime of the force controlled peeling appears, in our
experiments, to be only a transient phenomenon. This transient
behavior is clearly correlated to the increase in average peeling
velocity that is observed during the fall of the mass. It is
likely that the velocity will reach some elevated value on branch
III of figure \ref{SchemaFV}, but this evolution can not be
observed with the present setup. This result is in contradiction
with the existence of a stable stick-slip branch with a constant
average velocity as suggested in \cite{BARQU1997}. This
inconsistency is presently not explained and could be due to a
difference in the moment of inertia of the roller. We can
highlight that in \cite{BARQU1997} the presence of stick-slip in
the experiments corresponding to branch B of figure \ref{GrafVP}
was only inferred from the presence of acoustic emissions. In
contrast, the present experiment allows to access directly the
peeling point motion and to resolve the appearence and
disappearence of stick-slip motion.

\subsection{In the roller reference frame}\label{peelrefrol}

In sections \ref{rotreflab} and \ref{peelreflab}, we have
extracted the peeling point position $\ell_{\alpha}$ and the
position of the roller $\ell_{\beta}$ in the laboratory reference
frame. We are now able to determine the peeling point position in
the roller reference frame
$\ell_{\gamma}=\ell_{\alpha}-\ell_{\beta}$ which is the most
relevant variable of the problem.

\subsubsection{Triggered case}

\begin{figure}[h!]
\psfrag{X}[][][0.9]{$t$ (s)}
\psfrag{Y}[c][][0.9]{$|\ell_{\gamma}|$ (m)}
\psfrag{Z}[c][][0.9]{$\langle |\dot{\ell}_{\gamma}|
\rangle_{\small{cycle}}$ (m.s$^{-1}$)} \psfrag{A}[][][0.9]{(a)}
\psfrag{B}[][][0.9]{(b)}  \psfrag{C}[][][0.9]{(c)}
\psfrag{D}[][][0.9]{(d)}\centerline{
    \includegraphics[width=7.05cm]{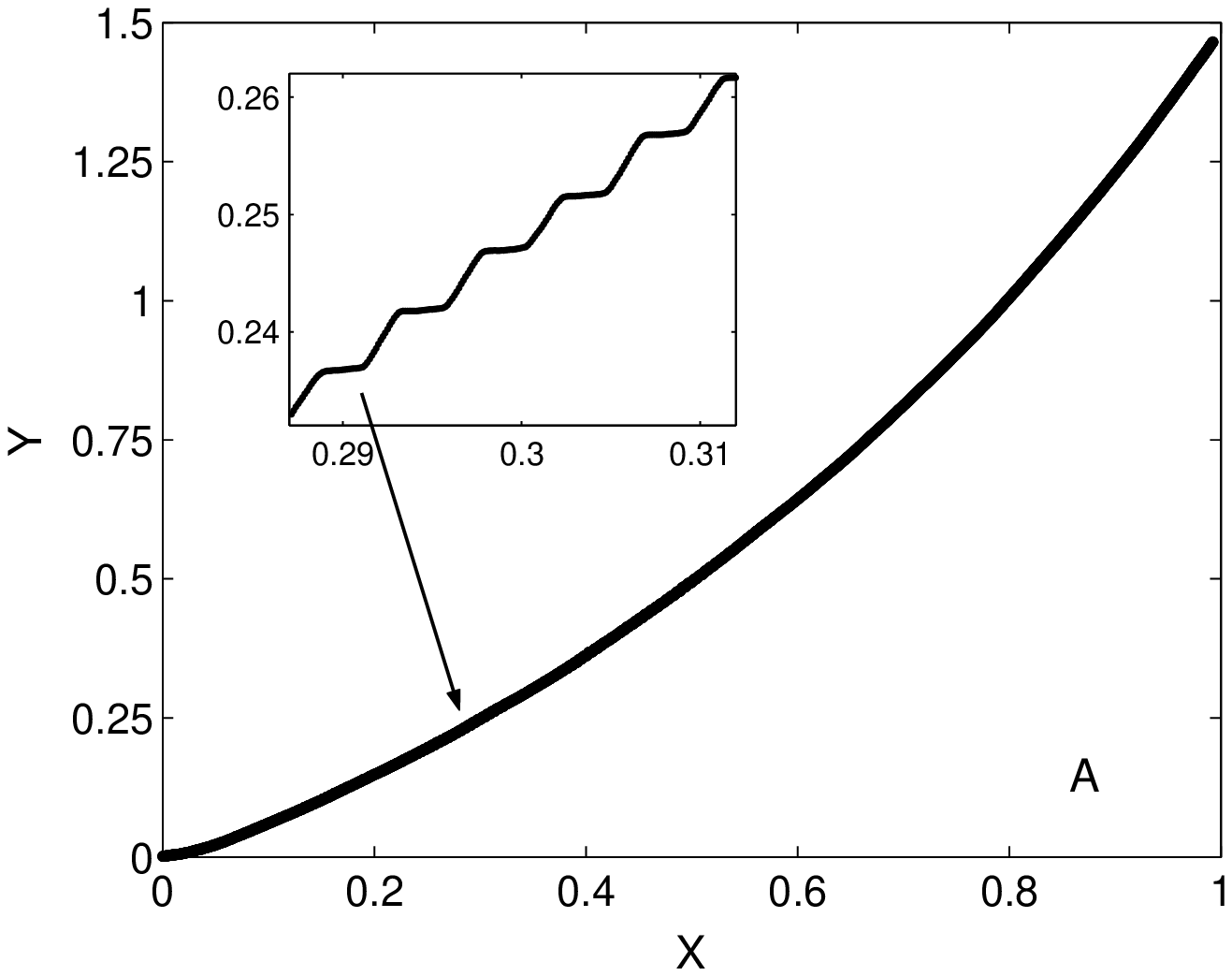}
    \includegraphics[width=7cm]{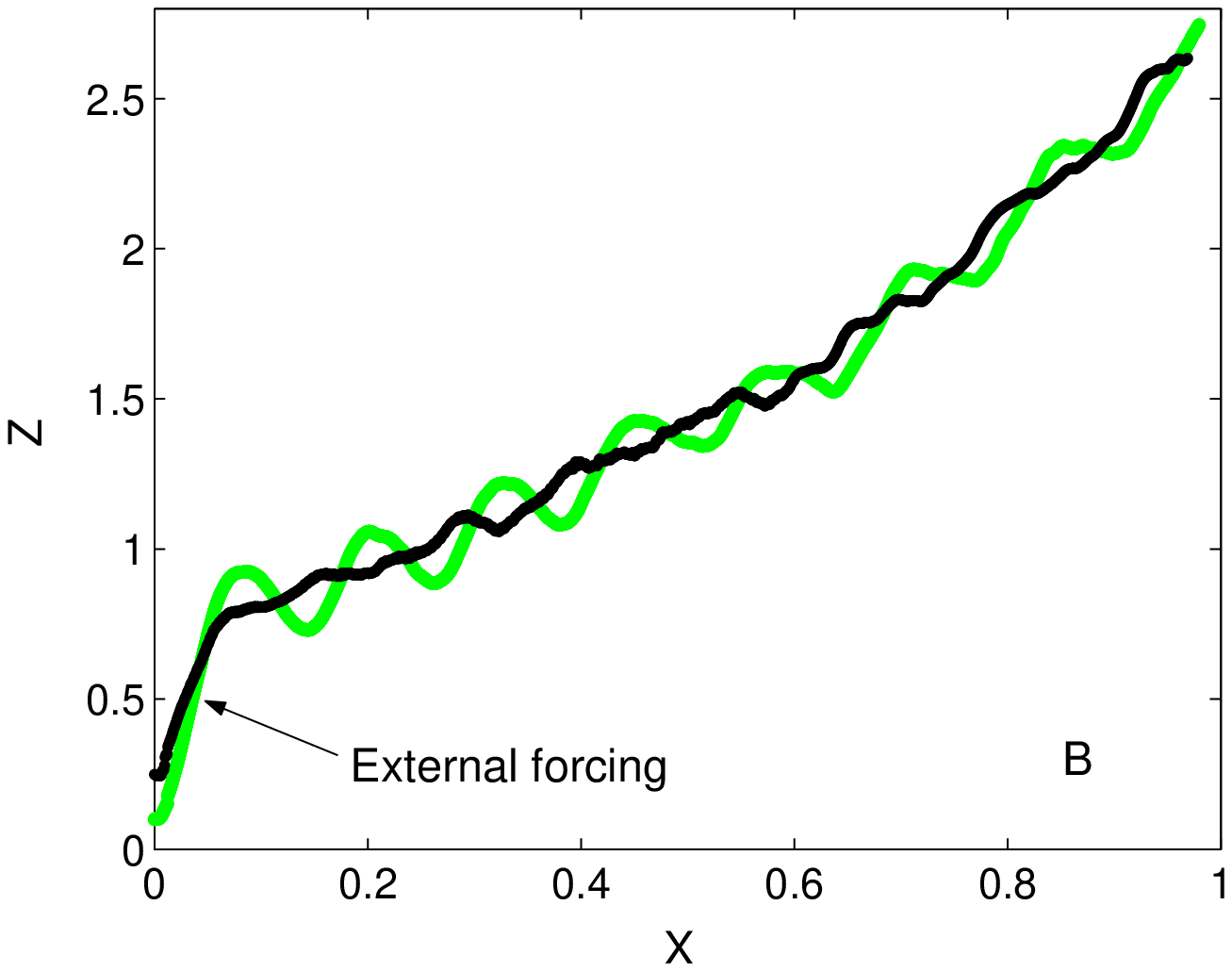}}
\vspace{0.2cm} \centerline{
    \includegraphics[width=7cm]{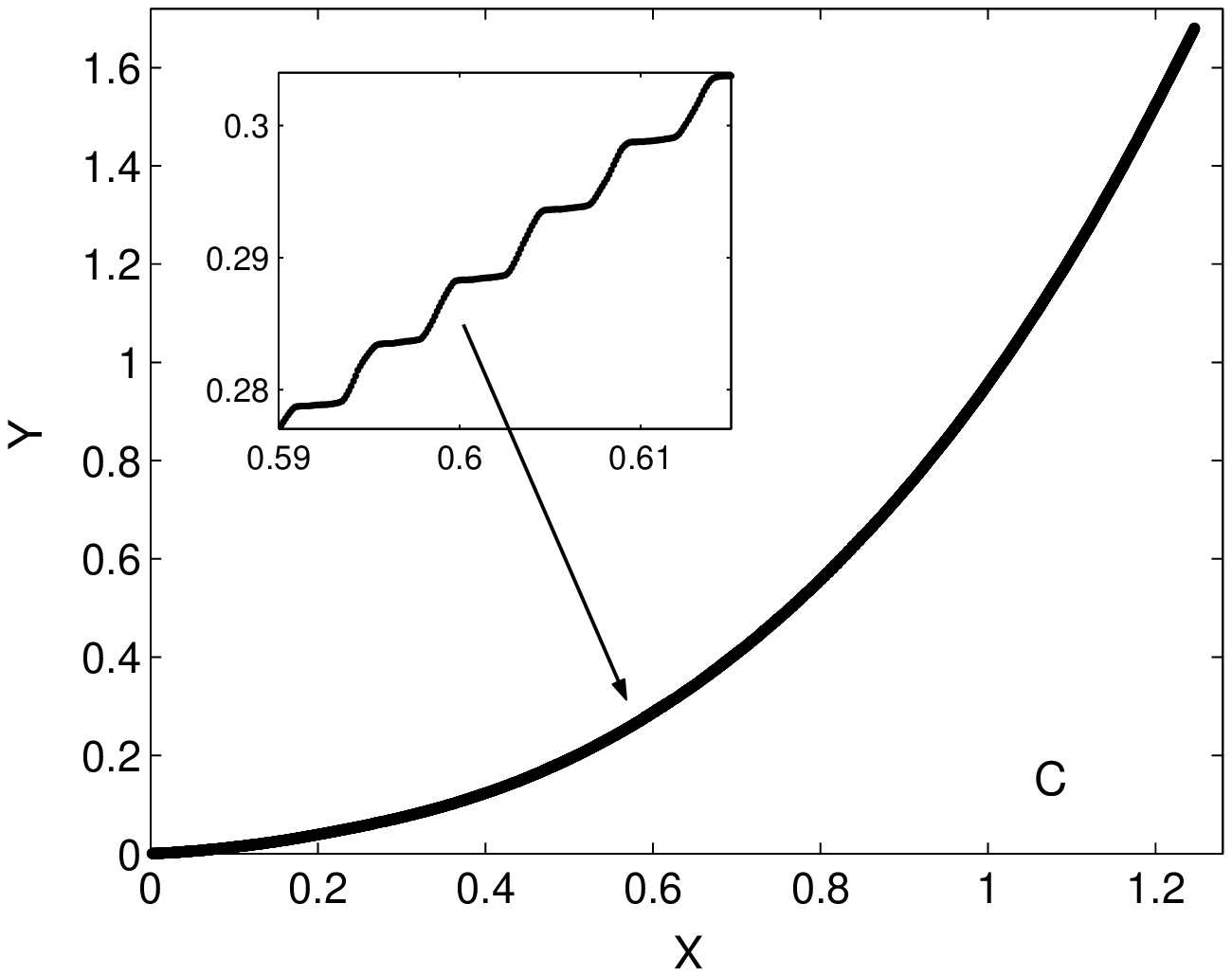}
    \includegraphics[width=7.1cm]{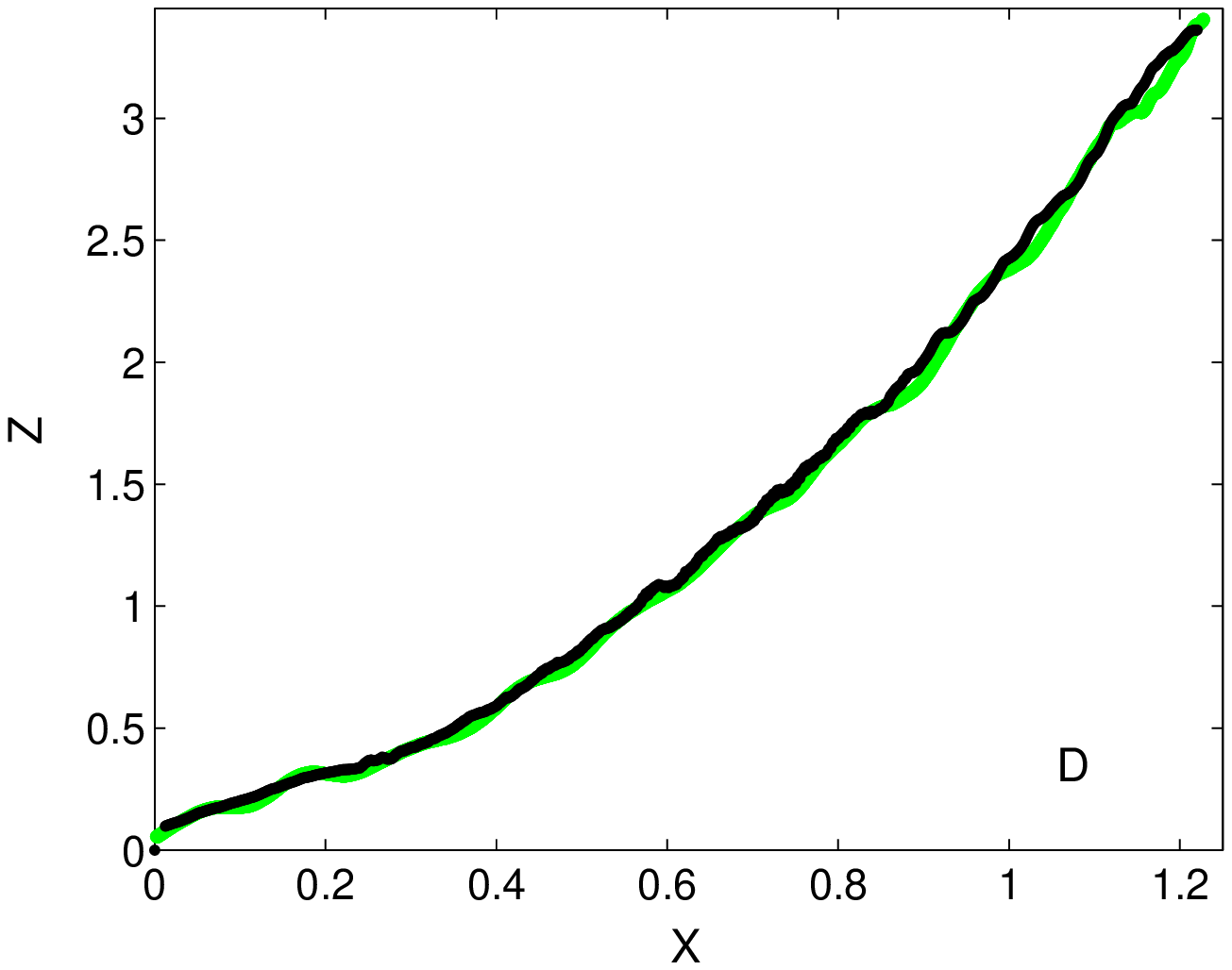}}
    \caption{\label{posvitrefrot} (a) and (c), absolute value of the peeling
    point position in the roller reference frame $\ell_{\gamma}$ as a function of time.
    The insets are zooms of these curves in a zone where stick-slip is observed.
    (b) and (d), corresponding (respectively to (a) and (c)) mean velocity of the peeling point (averaged over a stick-slip cycle)
    $\langle |\dot{\ell_{\gamma}}| \rangle_{cycle}$ as a function
    of time. The light grey (green) curves correspond to the velocity of the roller in the
    laboratory reference frame $\dot{\ell_{\beta}}$. (a) and (b) curves correspond to a triggered stick-slip peeling experiment
    performed with $m=195$g. (c) and (d) curves correspond to a spontaneous stick-slip peeling experiment
    performed with $m=245$g.}
\end{figure}

In figure \ref{posvitrefrot}(a) and (b), we show an example of
position and corresponding velocity of the peeling point in the
roller reference frame in the triggered case. We note that there
is an abrupt increase in the initial velocity that is the result
of the external forcing needed to trigger stick-slip. We also see
that the low frequency oscillations observed on the rotation
velocity of the roller have almost disappeared on the velocity of
the peeling point in the roller reference frame (cf. figure
\ref{posvitrefrot}(b)).

\subsubsection{Spontaneous case} In figures \ref{posvitrefrot}(c) and (d), we
show the position and corresponding velocity of the peeling point
in the roller reference frame in the spontaneous stick-slip case.
In contrast with the triggered case, the initial velocity
increases smoothly from zero. Similarly, the oscillations observed
in the laboratory reference frame have almost disappeared in the
roller reference frame.

\subsection{Qualitative evolution of the peeling point dynamics}

\begin{figure}[h!]
\psfrag{X}[][][0.9]{$t$-\scriptsize{offset} \normalsize{(s)}}
\psfrag{Y}[c][][0.9]{$\ell_{\alpha}$+\scriptsize{offset}
\normalsize{(m)}}
\psfrag{Z}[l][][0.9]{$|\ell_{\gamma}|$+\scriptsize{offset}
\normalsize{(m)}} \psfrag{A}[][][0.9]{(a)}
\psfrag{B}[][][0.9]{(b)}  \centerline{
    \includegraphics[width=12cm]{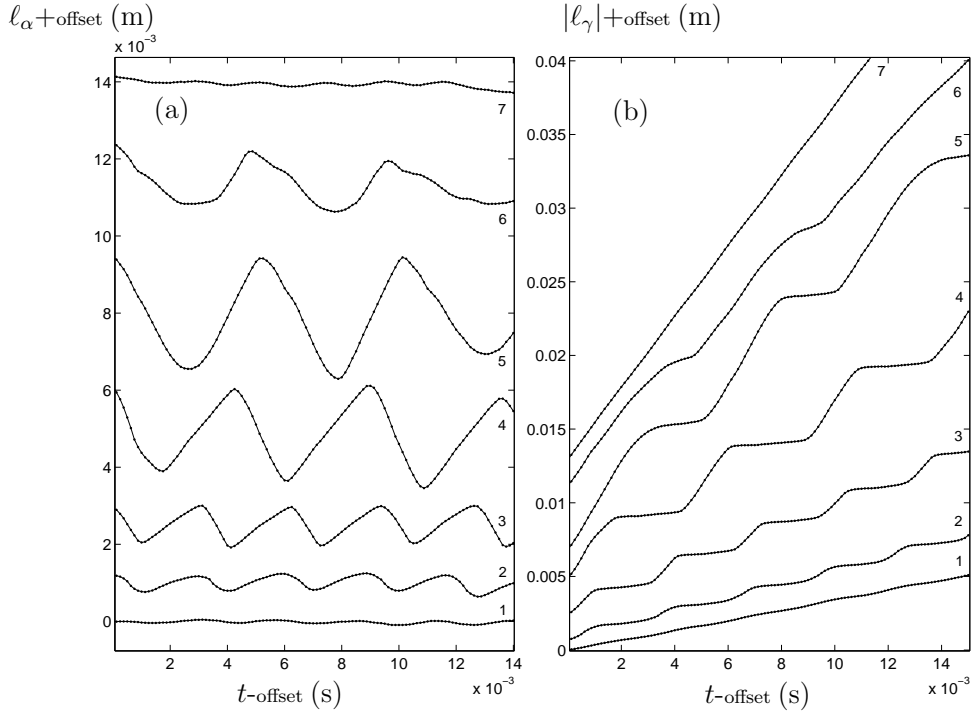}}
    \caption{\label{ss_lab_rol2450_02} (a) Peeling point position in the laboratory reference
    frame at different moments (time is increasing with the item number)
    of a spontaneous stick-slip experiment performed at $m=245$g. (b) Same data in the
    roller reference frame.}
\end{figure}

On figure \ref{ss_lab_rol2450_02}, one can see the peeling point
position in the laboratory reference frame, $\ell_{\alpha}$ (cf.
figure \ref{ss_lab_rol2450_02}(a)), and in the roller reference
frame, $|\ell_{\gamma}|$ (cf. figure \ref{ss_lab_rol2450_02}(b))
at different moments of a spontaneous stick-slip experiment
performed at $m=245$g. These figures show qualitatively the
changes in the stick-slip characteristics during the experiment.
The stick-slip amplitude and duration increase quite slowly with
time (curves $1$ to $5$) before abruptly decreasing back (curves
$5$ to $7$). We also see that the shape of the stick-slip cycle is
changing. A more quantitative study of the evolution of the
amplitude and duration as a function of the average peeling
velocity is presented in section \ref{statistics}.

On figure \ref{ss_lab_rol2450_02}(b), we see that the sharpness of
the transition between the stick and slip phases evolves during
the experiment as the peeling velocity increases. Actually, the
transition is smooth when the stick-slip just starts to appear at
low velocity or when it is close to disappear at high velocity. In
the intermediate velocity region, the stick-slip transition tends
to be sharper. We also note that, when the stick-slip regime is
close to disappear, the slip to stick transition tends to become
much smoother than the stick to slip transition.

\section{\label{statistics}The stick-slip average properties as a function of the mean peeling velocity}

In this section, we present results that have been extracted from
the two previously presented experiments performed with $m=195$
and $245$g. However, the behavior that will rise from these
experiments is reproducible from one experiment to the other for
the same experimental conditions.

\subsection{Stick and slip velocities}
\subsubsection{Spontaneous case}
\begin{figure}[h!]
\psfrag{A}[][][0.9]{(a)} \psfrag{B}[][][0.9]{(b)}
\psfrag{X}[c][][0.9]{$\langle
\dot{\ell_{\gamma}}\rangle_{\small{cycle}}$ (m.s$^{-1}$)}
\psfrag{Y}[c][][0.9]{$|\dot{\ell_{\gamma}}|$ (m.s$^{-1}$)}
\psfrag{Z}[c][][0.9]{$t$ (s)}\psfrag{C}[][][0.9]{(c)}
\psfrag{D}[][][0.9]{(d)} \centerline{
    \includegraphics[width=7cm]{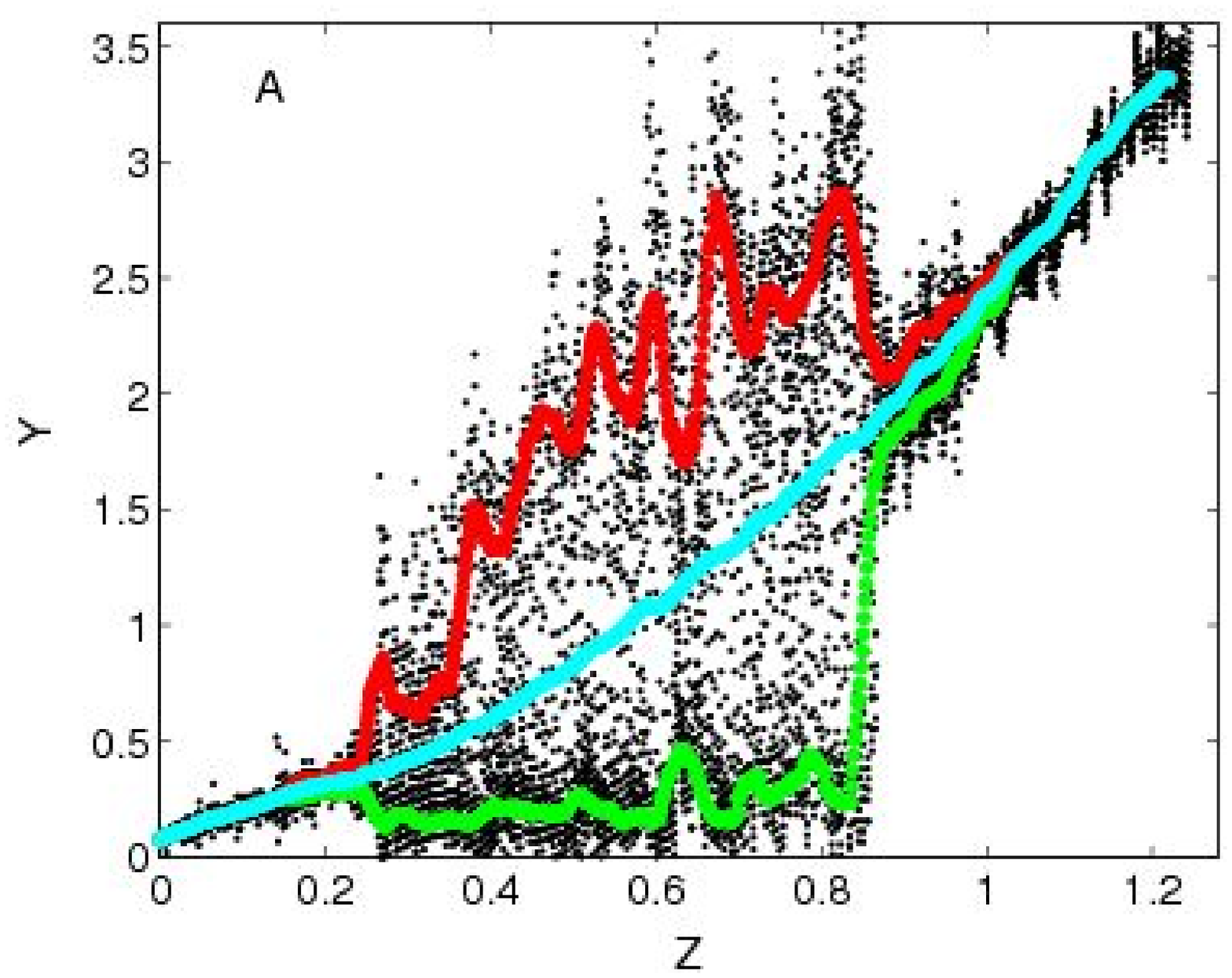}
    \includegraphics[width=7.05cm]{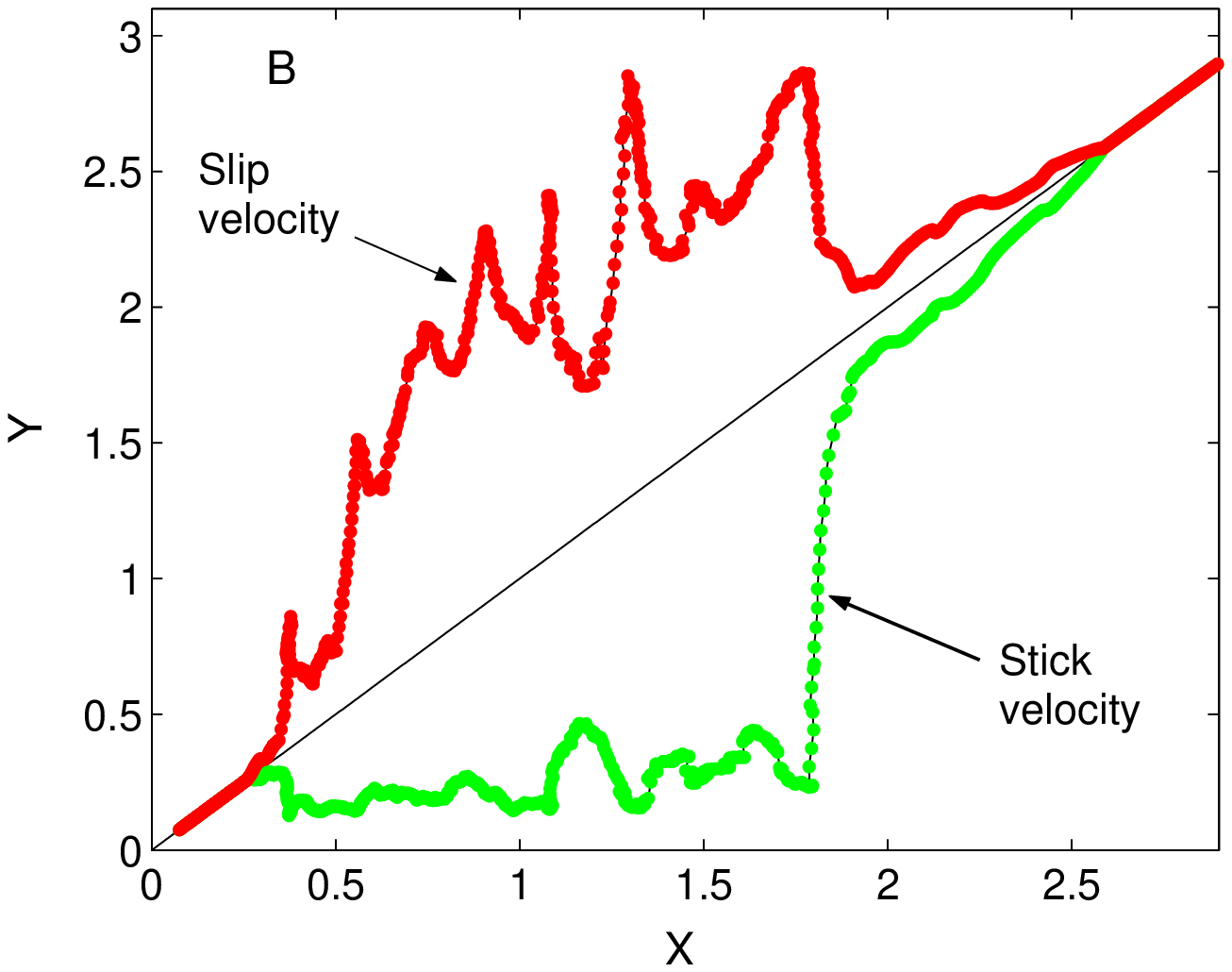}}
\vspace{0.2cm}
\centerline{
    \includegraphics[width=7cm]{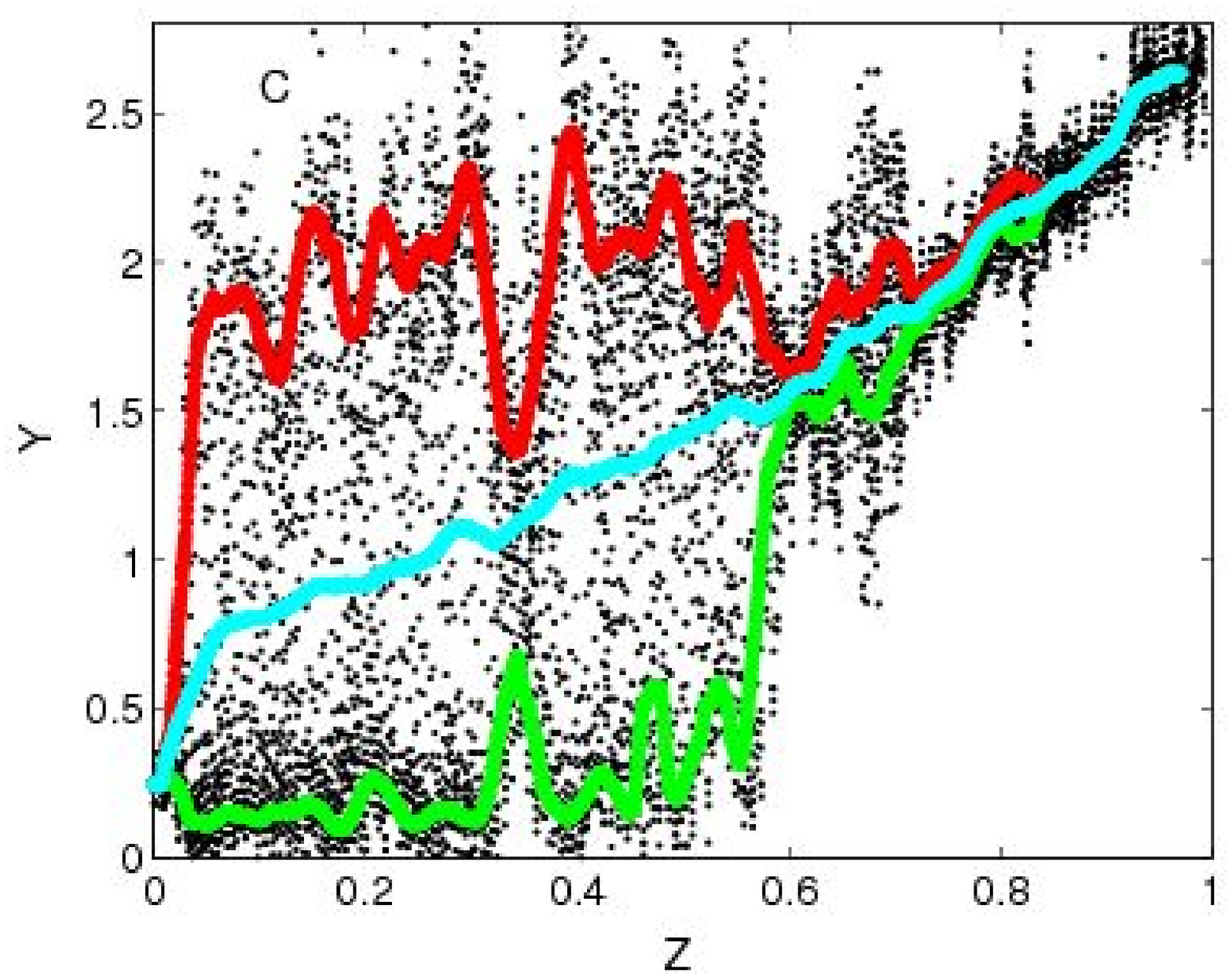}
    \includegraphics[width=7.05cm]{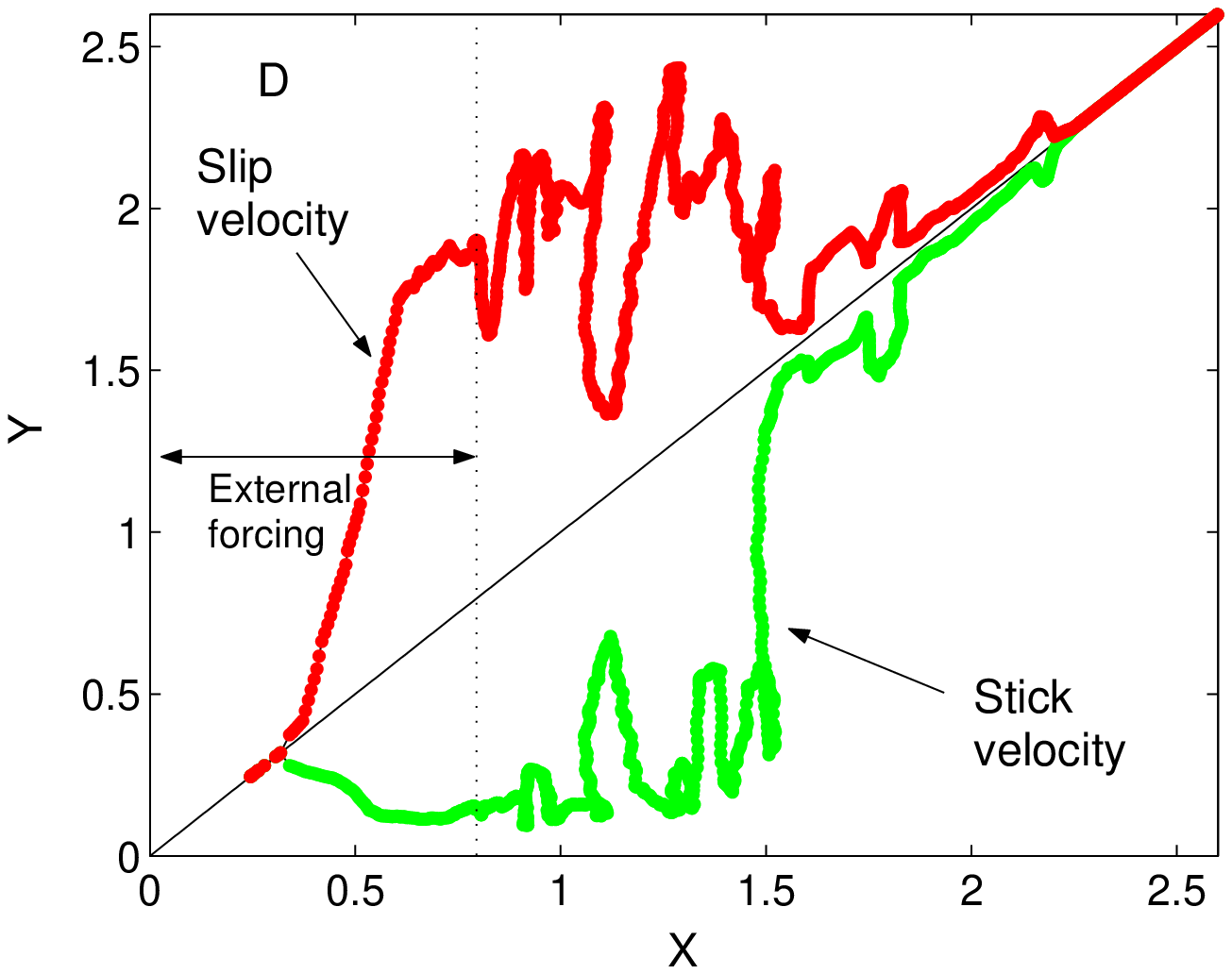}}
    \caption{\label{vssvit245} (a) and (c), instantaneous peeling velocity $|\dot{\ell_{\gamma}}|$ (black dots),
    average peeling velocity $\langle \dot{\ell_{\gamma}}\rangle_{\small{cycle}}$ (middle curve),
    average stick (bottom curve) and
    slip (top curve) velocities as a function of time. (b) and (d)
    corresponding (respectively to (a) and (c))
    average stick and slip velocities as a function of the average
    peeling velocity. (a) and (b) curves correspond to a spontaneous stick-slip peeling experiment
    performed with $m=245$g. (c) and (d) curves correspond to a triggered stick-slip peeling experiment
    performed with $m=195$g.}
\end{figure}

From the position signal of the peeling point in the roller
reference frame, we compute the instantaneous velocity and plot it
as a function of time (black dots in figure \ref{vssvit245}(a)).
The observed large velocity fluctuations are a signature of the
stick-slip motion. We extract from this instantaneous velocity the
mean stick and slip velocities averaged over a few stick-slip
cycles (cf. figures \ref{vssvit245}(a) and (b)). The stick-slip
motion initiates at a peeling velocity of about $0.25
\rm{m.s}^{-1}$ with average stick and slip velocities starting to
deviate from the average peeling velocity. This velocity
corresponds well to the maximum velocity reachable in a stable
peeling experiment $(0.20\pm 0.03)\rm{m.s}^{-1}$. When the average
velocity increases further, the stick velocity remains quite
stable with a value of $0.2-0.3 \rm{m.s}^{-1}$. It is important to
note that the average stick velocity remains close to the peeling
velocity just before the transition towards stick-slip. In
contrast, the slip velocity increases gradually from about $0.25
\rm{m.s}^{-1}$ up to $2.6 \rm{m.s}^{-1}$. The fluctuations that
are observed on the slip velocity, and in a lesser extent on the
stick velocity, are correlated to the low frequency oscillation of
the average peeling position. This shows that there is a
dependence of the stick-slip properties on the angle $\alpha$. The
stick-slip reduces in amplitude for an average peeling velocity of
$1.8$m.s$^{-1}$ and finally totally disappears for an average
peeling velocity of $2.6\rm{m.s}^{-1}$.

In the triggered case, the external forcing brings the average
peeling velocity to a value close to $0.8\rm{m.s}^{-1}$. The
stick-slip motion is initiated during this loading phase. For
average velocities larger than $0.8\rm{m.s}^{-1}$, both the stick
and slip velocities remain rather stable, despite some
fluctuations, until the stick-slip reduces its amplitude and
finally disappears (cf. figures \ref{vssvit245}(c) and (d)). By
comparing figures \ref{vssvit245}(b) and \ref{vssvit245}(d) we can
see a similar behavior for similar values of the average peeling
velocity.

\subsection{Stick-slip periods and amplitudes}
\begin{figure}
\psfrag{X}[c][][0.9]{$\langle
\dot{\ell_{\gamma}}\rangle_{\small{cycle}}$ (m.s$^{-1}$)}
\psfrag{Y}[c][][0.9]{$dt_{ss}$ (s)} \psfrag{Z}[c][][0.9]{Amplitude
(m)}\psfrag{A}[][][0.9]{(a)} \psfrag{B}[][][0.9]{(b)} \centerline{
    \includegraphics[width=7cm]{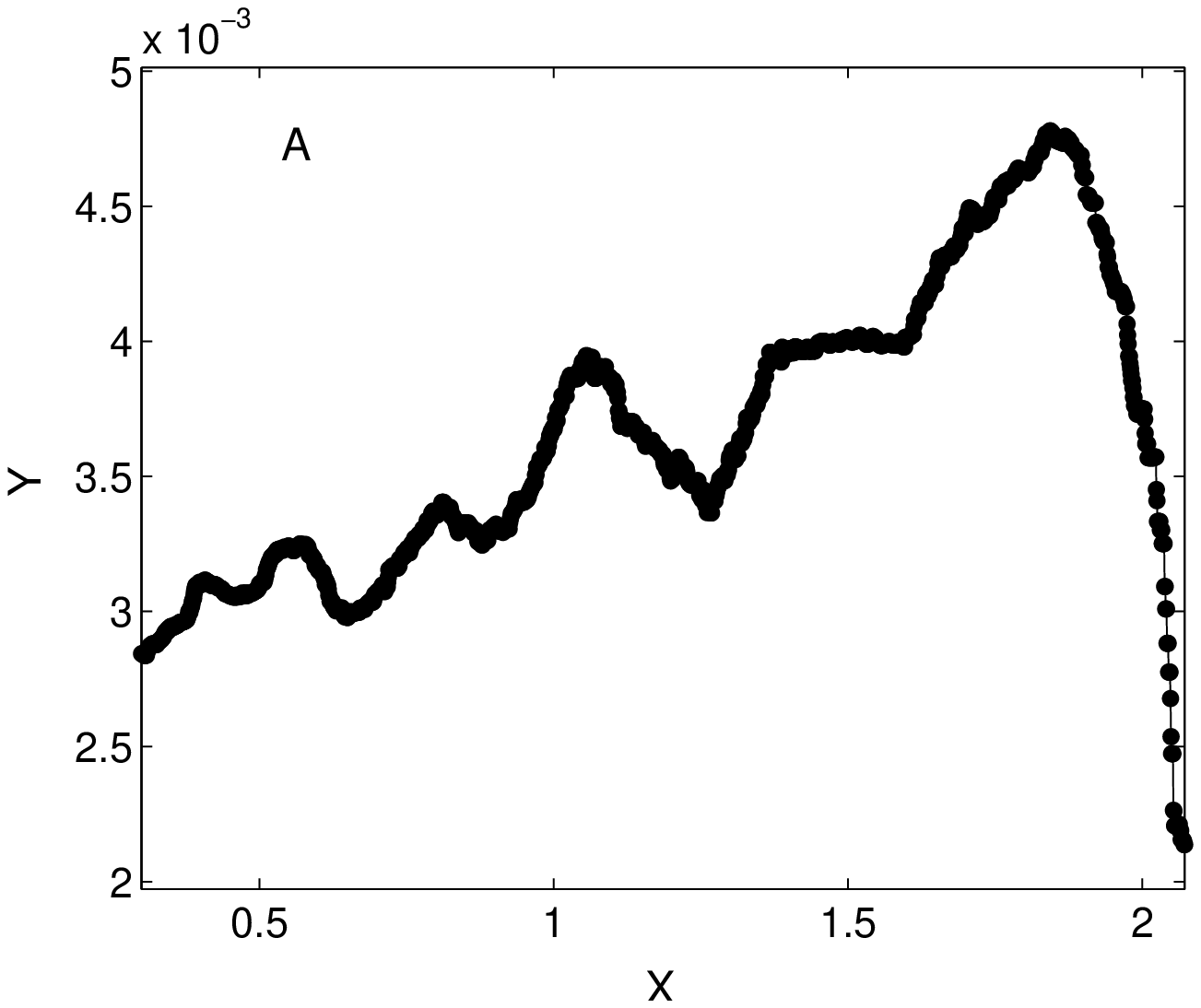}
    \includegraphics[width=7cm]{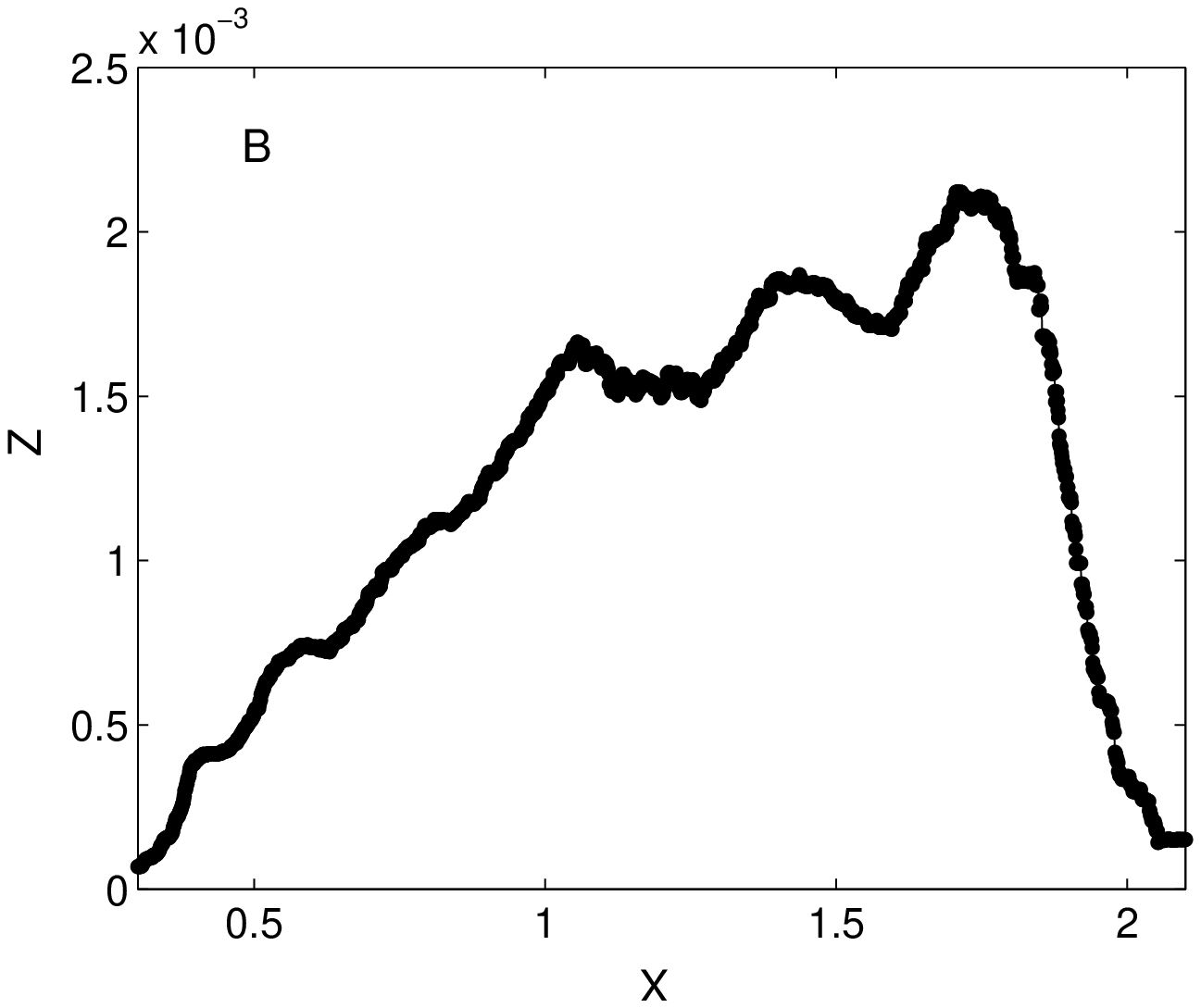}}
    \caption{\label{dtampvit245} (a) Stick-slip cycle duration and (b) amplitude
    (in the laboratory reference frame) as a function of the
    average peeling point velocity for a stick-slip peeling experiment performed with $m=245$g.}
\end{figure}

In figure \ref{dtampvit245}, we show the period and amplitude of
stick-slip as a function of the peeling velocity for a typical
experiment in the spontaneous case. We observe that the period
range is $2-5$ms and the amplitude range is $0-2$mm. Both the
period and amplitude are increasing with the peeling velocity.
When reaching a peeling velocity of about $1.8$m.s$^{-1}$, the
strong observed reduction in the stick-slip amplitude (down to
zero) is associated with a reduction of the stick-slip period.

\subsection{Relative stick and slip durations}
\begin{figure}[h!]
\psfrag{X}[c][][0.9]{$\langle \dot{\ell_{\gamma}}\rangle_{cycle}$
(m.s$^{-1}$)}
\psfrag{Y}[c][][0.9]{$\frac{dt_{\small{sk/sp}}}{dt_{\small{sk+sp}}}$}
\psfrag{Z}[c][][0.9]{$\frac{dt_{\small{sk}}}{dt_{\small{sk+sp}}}$}
\psfrag{U}[c][][0.9]{$\frac{dt_{\small{sp}}}{dt_{\small{sk+sp}}}$}
\centerline{
    \includegraphics[width=7cm]{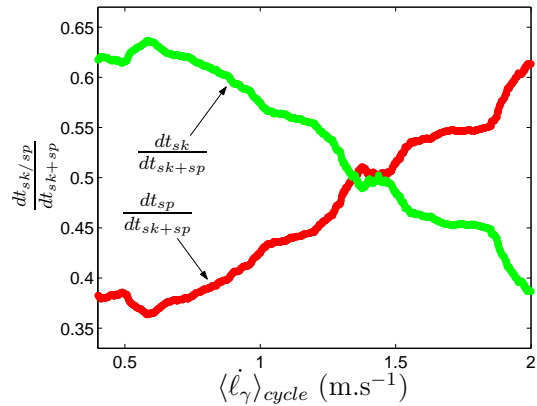}}
    \caption{\label{dtratio245} Ratio of the stick (light grey/green points) and slip (strong grey/red points) phases
    duration with the stick-slip duration as a function of the
    average peeling point velocity for a
    stick-slip peeling experiment performed with $m=245$g.}
\end{figure}

Another important feature of the stick-slip motion is that the
relative durations of the stick and slip phase are evolving with
the average peeling velocity. On figure \ref{dtratio245}, we show
the fraction of time during a stick-slip period spent in the stick
and in the slip phase. Remarkably, there is never more than a
factor of $2$ between the stick duration and the slip duration.
This observation rules out models where the slip duration was
assumed to be much shorter than the stick one. Even more
remarkably, the stick duration, initially about twice the slip
duration, gradually decreases with the peeling velocity and
becomes smaller than the slip duration, down to almost one half.

These results might look surprising compared to the usual
frictional stick-slip phenomenon. Indeed, in the case of the
traditional spring-block experiment stick-slip \cite{JOHAN1994,
BAUMB2006}, the stick phase is often expected to be much longer
compared to the slip phase. However, the slip phase duration can
become comparable to the stick phase one when the system is near
to the threshold of stick-slip appearance \cite{BAUMB2006,
BAUMB1994}. In contrast, for the stick-slip peeling of an adhesive
tape, this remains true far from the threshold.

\section{Discussion}

In this article, we have presented the results of a direct fast
imaging of the stick-slip peeling of an adhesive tape under force
controlled conditions. The motion of the roller and of the peeling
point are measured in the laboratory reference frame with an
excellent resolution in order to study at the same time the global
dynamics of the system and the details of the peeling dynamics. In
these experiments, whether the stick-slip is spontaneous or
triggered, we observed that it is only a transient phenomenon that
happens in conjunction with a progressive acceleration of the
falling load. The velocity at which the stick-slip appears
spontaneously is close to the limit velocity for the regular peeling
(about $0.2$m.s$^{-1}$). Then, for an increasing average peeling
velocity, the stick-slip amplitude grows before it abruptly reduces
at a velocity of about $1.8$m.s$^{-1}$, and finally disappears
completely at $2.6$m.s$^{-1}$.

In all our observations, the peeling was shown to be accompanied by
an oscillation in the roller velocity at about $10$Hz that can be
modelled in terms of the harmonic torque induced by the oscillations
of the peeling point angle. These oscillations have an influence on
the characteristics of the stick-slip such as period and amplitude.
The duration of the stick and slip phases reveals to be comparable
(ratio from 0.5 to 2), with a longer stick duration at low peeling
velocity and a longer slip duration at large peeling velocity. The
sharpness of the transition between the stick and slip phases
evolves during an experiment as the mean peeling velocity increases.
Actually, the transition is smoother when the stick-slip just starts
to appear at low velocity or when it is close to disappear at high
velocity. In the intermediate range of velocity, the stick-slip
transition tends to be sharper.

Our data are in contradiction with the existence of a stable
stick-slip branch with constant average velocity as suggested in
\cite{BARQU1997}. We can highlight that in \cite{BARQU1997} the
presence of stick-slip in the experiments for the branch B of figure
\ref{GrafVP} was only inferred from the presence of acoustic
emissions during the test. To overcome this contradiction, we plan
to acquire acoustic emissions in parallel to the imaging of the
peeling point motion. It is also important to check whether
increasing the falling height of the mass in our experiment would
help reaching a stationary peeling regime or not.

The measurement technique that we have developed is precise enough
to make quantitative comparison with theoretical models.  In order
to clarify the physics at stake in the peeling of an adhesive
tape, more experimental work will be necessary. For instance,
understanding the complex stick-slip statistics at high peeling
rate \cite{CICCO2004} will require studies in the velocity
controlled regime for which the stick-slip dynamics is stationary.

\ack We thank S. Ciliberto for insightful discussions.

\end{document}